\documentclass[a4paper,10pt]{article}
\usepackage{authblk}
\usepackage{graphicx}
\usepackage{bm}
\usepackage[colorlinks,linkcolor=blue, urlcolor=blue, anchorcolor=blue, citecolor=blue]{hyperref}
\usepackage{subfigure}
\usepackage[section]{placeins}
\usepackage{lipsum}
\begin{document}

\title{High-speed confined granular flows down inclines revisited}

\author{Yajuan Zhu}
\author{Renaud Delannay}
\author{Alexandre Valance}


\affil{
Institut de Physique de Rennes, 
CNRS UMR 6251, Univ Rennes, 35042 Rennes CEDEX, France
}

\maketitle
\begin{abstract}
Recent numerical work has shown that high-speed confined granular flows down inclines exhibit a rich variety of flow patterns,
including dense unidirectional flows, flows with longitudinal vortices and supported flows characterized by a dense core
surrounded by a dilute hot granular gas  \cite{Brodu2015}. 
Here, we revisit the results obtained in \cite{Brodu2015} and present  new features characterizing these flows. 
In particular, we provide vertical and transverse profiles for the packing fraction, velocity and granular temperature.
We also characterize carefully the transition
between the different flow regimes 
and show that the packing fraction and the vorticity can be successfully used to describe these transitions.
Additionally, we emphasize that the effective friction at the basal  and side walls can be described by a unique function of a dimensionless
number which is the analog of a Froude number: $Fr=V/\sqrt{gH\cos \theta}$ where $V$ is the particle velocity at the walls, 
$\theta$ is the inclination angle and $H$ the particle holdup (defined as the depth-integrated particle volume fraction).
This universal function bears some similarities with the $\mu(I)$ rheological curve derived for dense granular flows.
\end{abstract}


\section{Introduction}
The scientific community has paid particular attention
to gravitational granular flows over the past 20 years. These flows are ubiquitous in natural and industrial
processes \cite{Midi2004,Delannay2017}. However, their modeling and understanding still leave
us with open issues. The complexity arises from grain-grain interactions, but also from 
grain-boundary interactions which may induce correlations over distances much greater than a grain diameter. 

The inclined plane geometry was the most employed configuration to study gravity-driven granular flow \cite{Pouliquen1999,Forterre2008}.
It is simple and relevant for many practical situations, but it  
can be also seen  as a rheological test with constant friction. 
To date, experiments \cite{Pouliquen1999} and simulations \cite{Silbert2001} have focused mainly
on mildly sloping planes, leading to slow and dense flows which are now fairly well understood \cite{Midi2004,Delannay2007}.
More complex flows, including span-wise
vortices \cite{Forterre2001,Forterre2002,Borzsonyi2009}, were obtained at slightly higher angles suggesting 
that upon further steepening, granular flows may reveal original features.

Obtaining steady and fully developed (SFD) flows at steep angles is both a experimental and numerical challenge.
Indeed, for unconfined flows, there is in general a limit angle above which flows keep accelerating. This limit angle
may depend on many parameters such as the mechanical properties of the grain and  the nature of the base
(flat or bumpy). A simple way to obtain SFD flows at high angles is to introduce frictional side walls. Indeed,
if the grain–wall friction coefficient is high enough, one may expect
that the base friction supplemented by the sidewall friction will be able to balance
the driving component of the weight. This is what has been done recently by Brodu et al. \cite{Brodu2015}
by means of discrete element method simulations. These simulations showed that SFD flows can be produced at arbitrary high angles
and revealed the existence of new flow regimes characterized by complex internal structures with heterogeneous particle volume fraction and
secondary flows \cite{Brodu2013,Brodu2015,Ralaiarisoa2017}. 
One of these regimes, referred to as “supported flow”, is particularly interesting since
it displays uncommonly high bulk velocity, the granular flow being “supported” on a dilute  granular  gaseous layer of highly
agitated grains. Similarly to an air-cushion suspension, this layer reduces the effective wall friction and increases
significantly the bulk velocity. These "supported" flows are particularly interesting with respect to geophysical issues.
The reduction in the effective friction due to the gaseous granular layer could indeed explain unexpected
long run-out distances of large granular avalanches.

In this paper, we revisit the results obtained by Brodu et al \cite{Brodu2015} on high-speed confined granular flows and
present new features characterizing the supported flow
regime, including transverse profiles of particle volume fraction and velocity, and also, vertical and horizontal
profiles of particle velocity fluctuations. 
We also describe in detail the transition between the different flow regimes and provide a unified picture to describe
the variation of the effective friction at the boundaries, in terms of a Froude number defined as
 $Fr=V/\sqrt{gH\cos \theta}$ where $V$ is the particle velocity at the walls, 
$\theta$ is the inclination angle and $H$ the particle holdup (defined as the depth-integrated particle volume fraction).

The outline of the paper is the following. In section 2 we briefly present the flow geometry and the discrete element method used for the
simulations. Then, in section 3 we describe the different flow regimes according to the particle hold-up and inclination angle. 
Section 4 is devoted to the analysis of the transition between the different flow regimes. In section 5, we provide
a full characterization of the supported flow regime in terms of packing fraction, velocity and temperature fields. 
Basal and sidewall friction and their relation with velocities at the boundary are discussed in section 6. 
Finally, we conclude in section 7.

\section{Flow geometry}
\begin{figure}[htb]
\begin{center}
\includegraphics[width=.5\columnwidth]{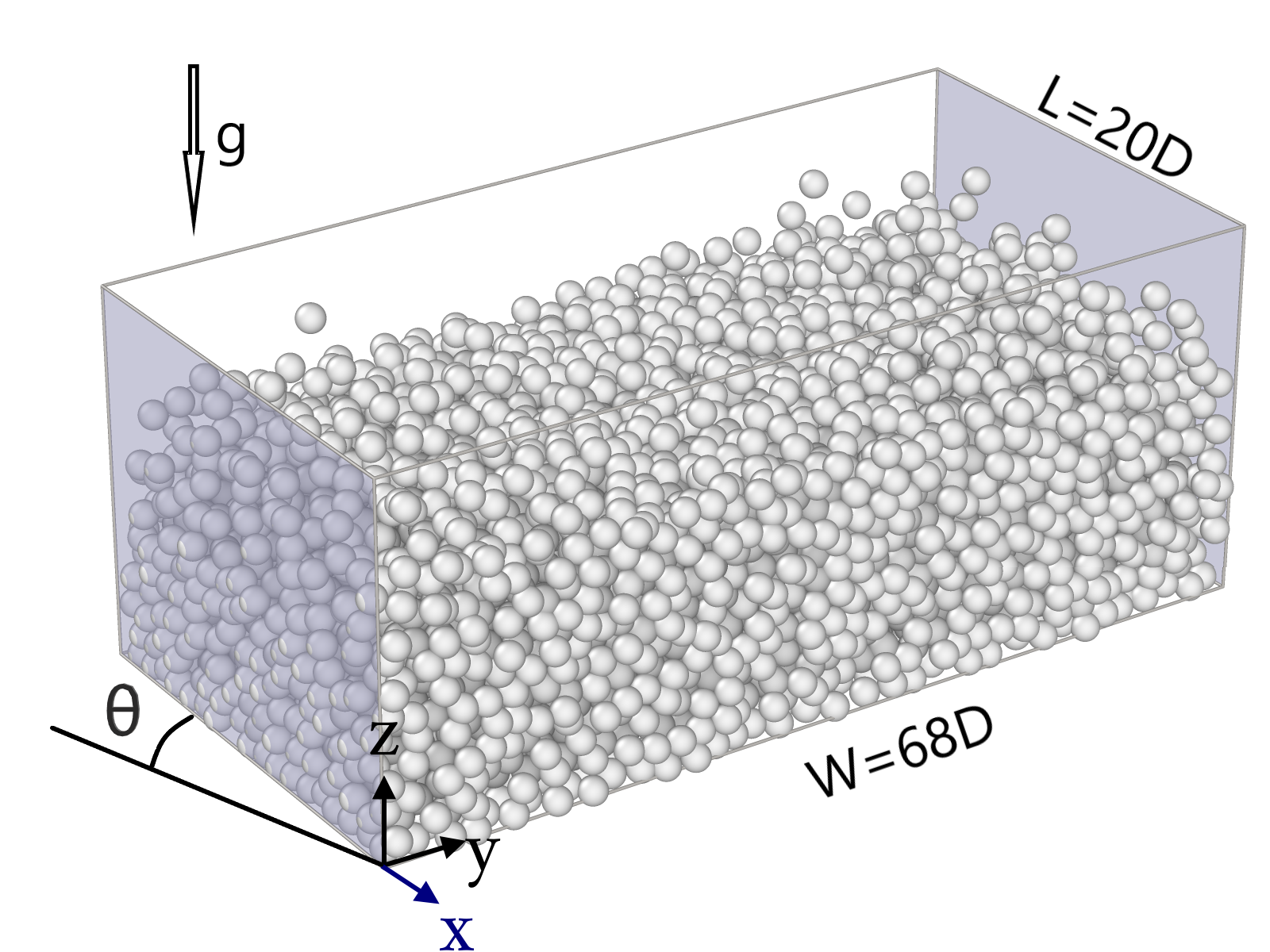}  
\end{center}
\caption{Scheme of the simulated system. The channel consists of frictional and flat bottom and sides and is inclined with an angle $\theta$
with respect to the horizontal. 
The longitudinal length $L$ and width $W$ of the channel are set to $20D$ and to $68D$, respectively.
The channel is not bounded in the $(0z)$ direction and we use periodic boundary conditions in the stream-wise direction.
}
\label{scheme}
\end{figure}

We consider gravity-driven chute flows with flat frictional bottom and sides, as shown in Fig.\ref{scheme}. 
The chute is inclined with an angle $\theta$ with respect to the horizontal. 
$(0x)$ is the main direction flow, $(0y)$ the cross-wise direction and $(0z)$ is the direction perpendicular to the flow base.
This geometry is similar to that used in \cite{Taberlet2003,Richard2008,Brodu2013,Brodu2015}. Here, the simulation
cell has similar dimensions as those employed
by Brodu and co-workers \cite{Brodu2013,Brodu2015}. In particular, the longitudinal length $L$ and the gap $W$ between the side-walls are
set to $L=20D$ and $W=68D$, respectively (where $D$ is the particle diameter).
The channel is not bounded in the $(0z)$ direction and periodic boundary conditions are employed in the stream-wise direction $(0,x)$.

We use soft-sphere molecular dynamics simulations where particles in contact can overlap \cite{Brodu2013,Brodu2015}.
The contact forces between two particles have both a normal and a tangential
component. The normal force, $F_n$, is  modeled by a spring and a dashpot: 
$F_n = k_n \delta  − \gamma_n \dot{\delta}$,  where $\delta$ is the overlap and its derivation with respect to time, respectively, and, $k_n$ and
$\gamma_n$ are the spring stiffness and the viscous damping coefficient, respectively.
A similar model is used for the tangential component enforced by the Coulomb friction $|Ft| <\mu|Fn|$ where $\mu$ is
the friction coefficient.

We employ the same mechanical parameters as those in the experiments by Louge et al. \cite{Louge2001} and in the numerical
simulations of Brodu and co-workers \cite{Brodu2013,Brodu2015}. 
We choose values for $k_n$ and $\gamma_n$ (resp. $k_t$ and $\gamma_t$) such that the normal restitution coefficient $e_g^n$ 
(resp. the tangential one $e_g^t$) is equal to $e_g^n=0.972$ (resp. $e_g^t=0.25$).
The particle-particle friction coefficient is set  $\mu_g=0.33$. 

The walls (i.e., the bottom and the side-walls) are treated like spheres of infinite
mass and radius. The normal restitution coefficient $e_w^n$ and the friction coefficient $\mu_w$ for the
grain-wall interaction are set to $e_w^n=0.8$ and $\mu_w=0.593$, respectively. These values are also taken from Louge's experiments \cite{Louge2001}.

The control parameters of the simulation are the mass holdup $H$ and the inclination angle $\theta$, while the
the channel width $W$ is kept fixed (i.e., $W=68D$).
The particle hold-up $H$, defined as the depth-integrated particle volume fraction (i.e., $H=\int_0^\infty \phi(z) dz$, where $\phi$ is the 
particle volume fraction at height $z$) is varied from $4D$ to $12D$, and the inclination from $15^\circ$ to $50^\circ$.

In the following, unless otherwise specified, particle volume fraction, velocity and velocity fluctuations are averaged 
spatially in the stream-wise direction
and over time during $30$ time units (i.e., $\sqrt{D/g}$).  

\section{Steady and Fully developed flow regimes}

In \cite{Brodu2015}, 5 different flow regimes were identified: (i) A  unidirectional, dense and layered flow  
(labeled here after U for unidirectional); (ii) A dense and layered flow regime with two small longitudinal vortices
located at the side wall and close to the free surface (named SR for side rolls); (iii) a roll regime  which exhibits a pair of
counter-rotative longitudinal vortices that spans the entire width of the cell; (iv) and (v) two types of unusual flows 
characterized by a dense core floating over a dilute basal layer (referred here after to as supported regimes).
\begin{figure}[hbt]
\begin{center}
\subfigure[]{\includegraphics[width= 0.3\columnwidth ]{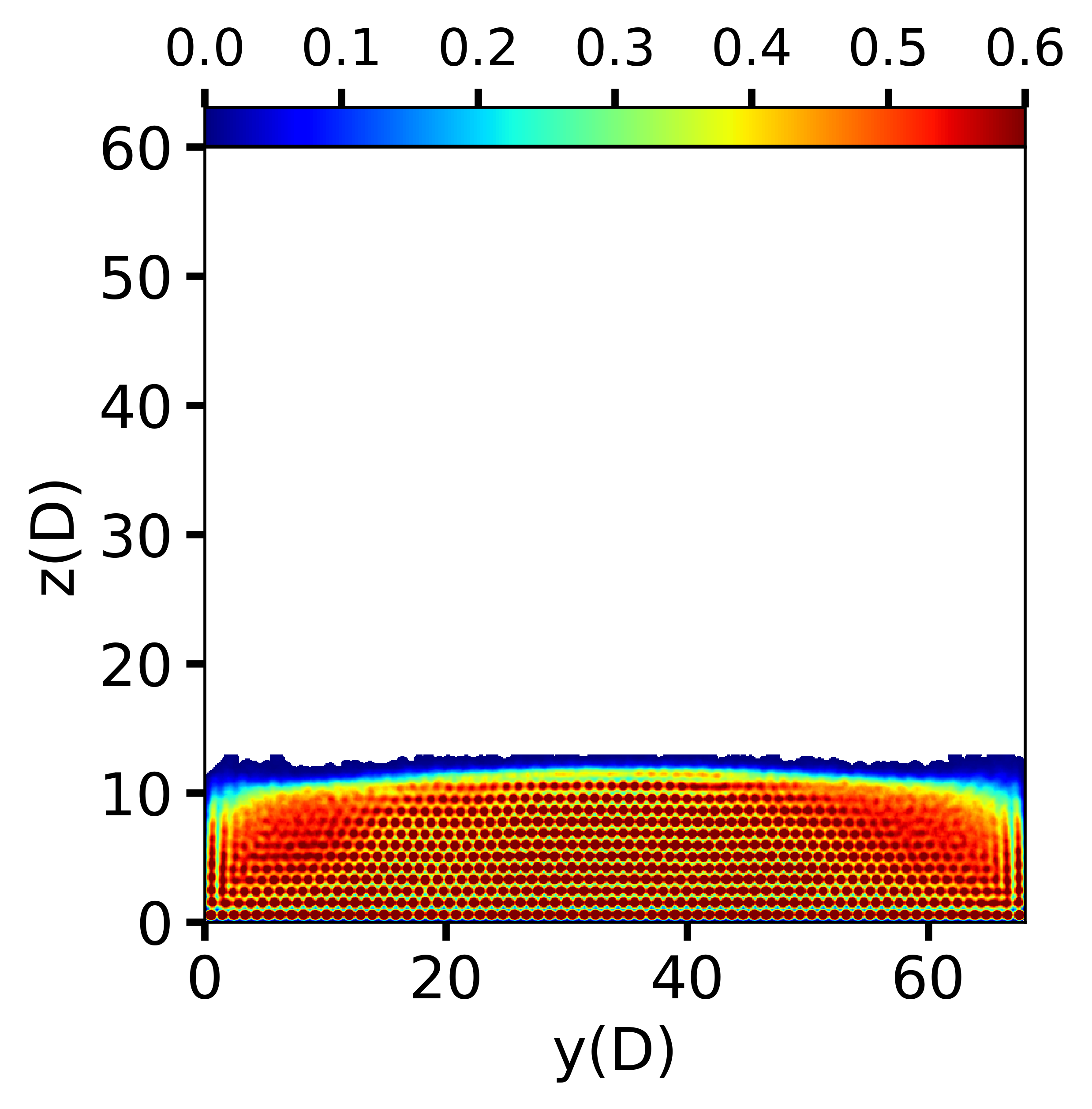}}
\subfigure[]{\includegraphics[width= 0.3\columnwidth ]{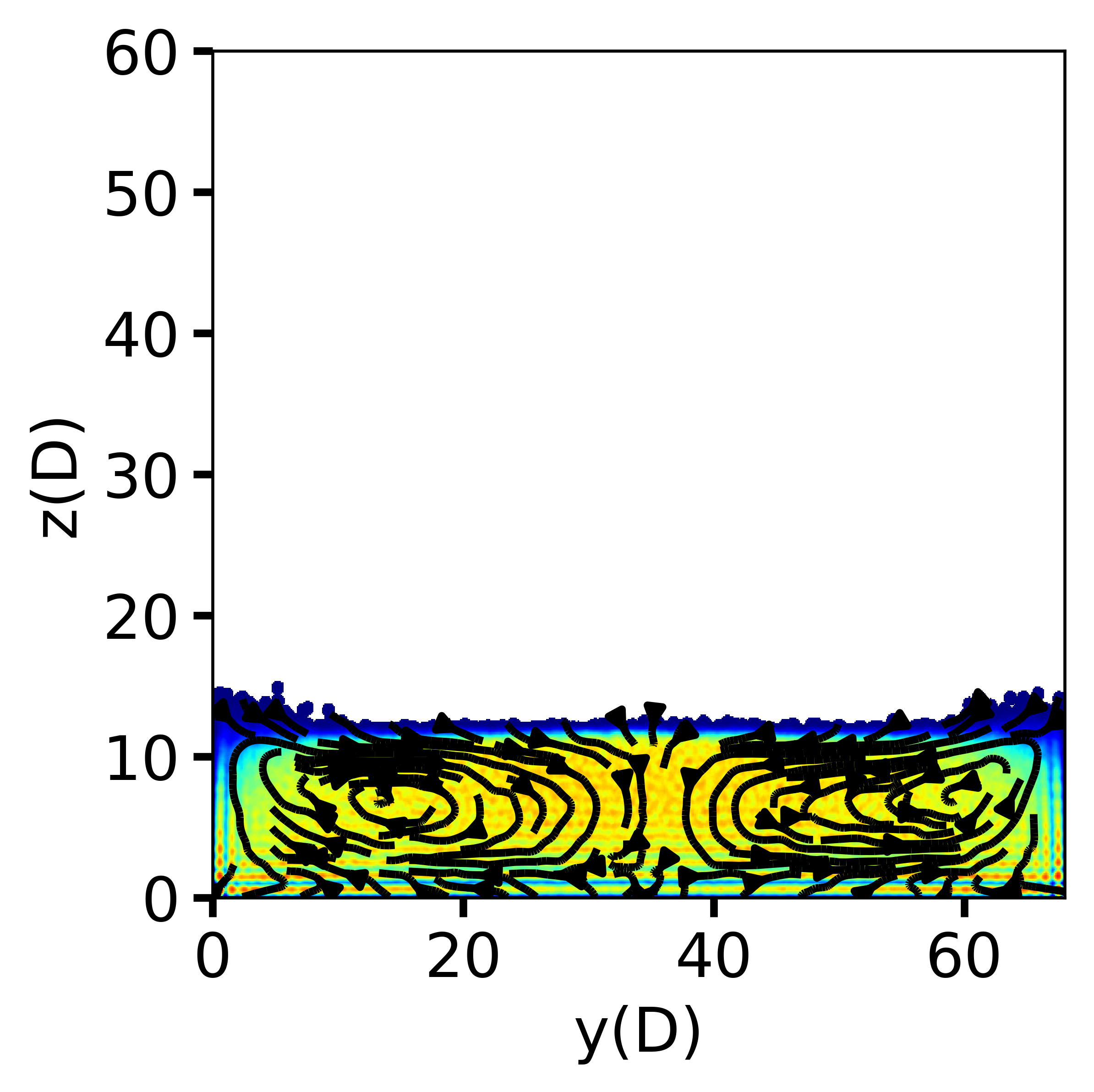}}\\
\subfigure[]{\includegraphics[width= 0.3\columnwidth ]{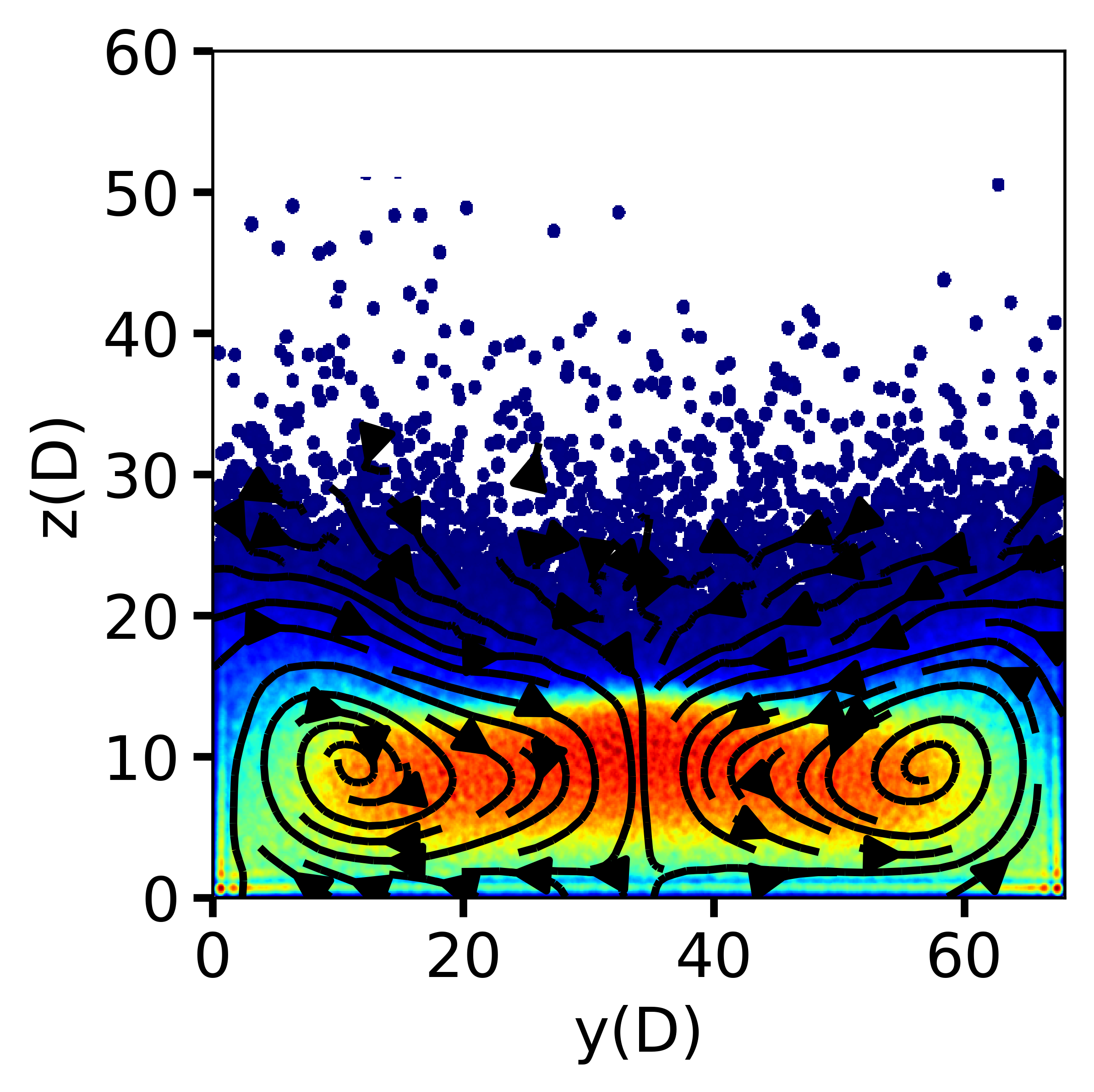}}
\subfigure[]{\includegraphics[width= 0.3\columnwidth ]{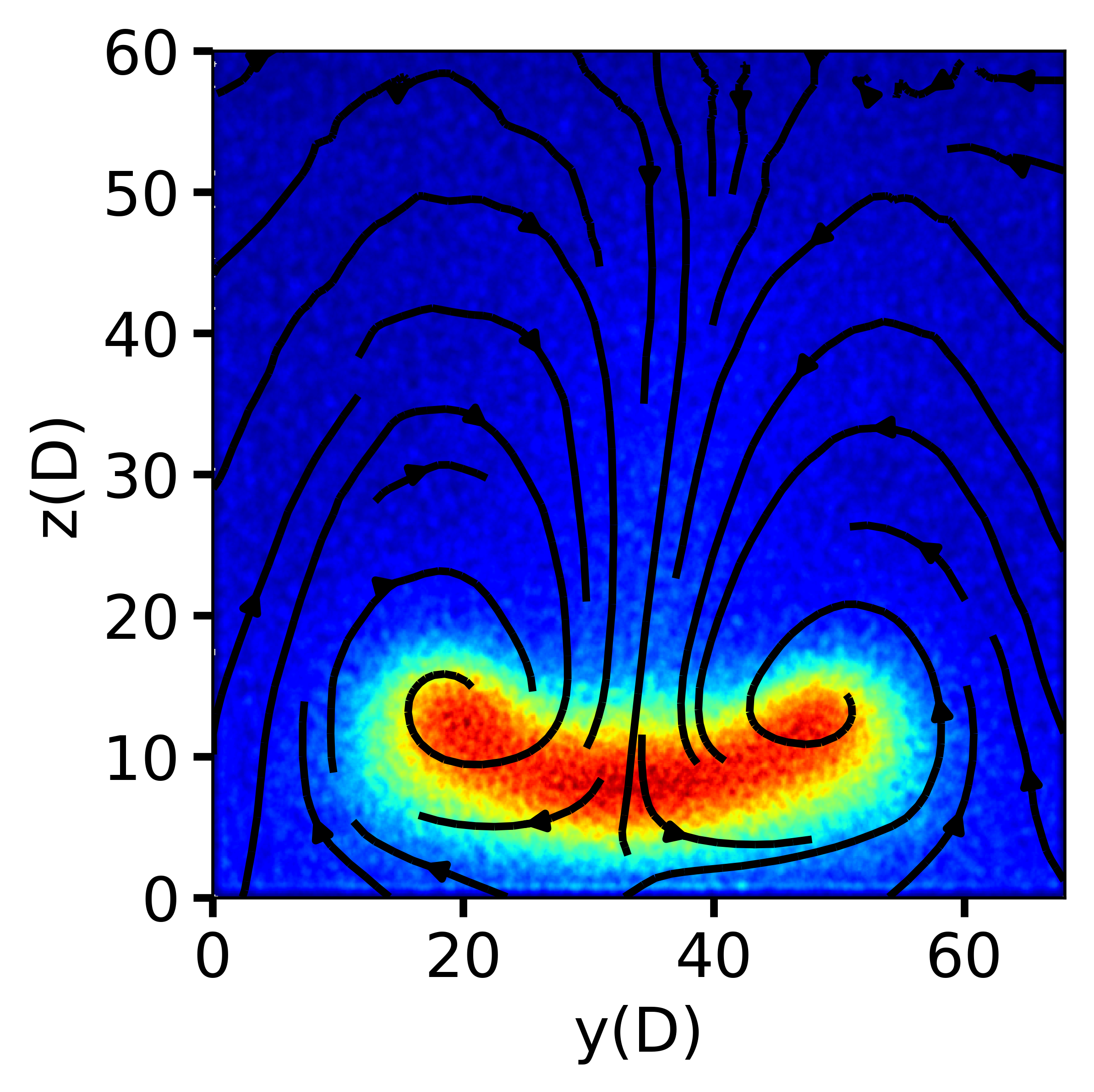}}
\end{center}
\caption{ Volume fraction map in the cross-section of the flow for different flow regimes with a fixed particle hold-up ($H=6D$).
The color codes the volume particle fraction (blue indicates dilute regions while red dense ones) and the solid lines stands for the streamlines.
(a) Unidirectional, dense and layered  flow ($U$) ($\theta=19^\circ$); (b) roll regime ($R^-$) ($\theta=20^\circ$); 
(c) and (d) supported flow regimes with a symmetric core ($\theta=27^\circ$) and an asymmetric core  ($\theta=40^\circ$), respectively. }
\label{compa_map}
\end{figure}
Four of these regimes are  illustrated in Fig.~\ref{compa_map} for $H=6D$
where the two-dimensional particle volume fraction map in the cross-section of the flow are presented together with the streamlines. 

Additional features are worth mentioning. While the unidirectional flow presents a uniform particle volume fraction through the depth,
the roll regime exhibits a slight particle density inversion, that is a lower particle volume fraction close to the bottom and
a higher volume fraction in the bulk flow. The apparition of the longitudinal rolls can be explained as the
result of a "Rayleigh-B\'enard"-like instability \cite{Forterre2002}. This roll regime has been observed
in discrete numerical simulations for the first time for unconfined geometries \cite{Borzsonyi2009} (i.e., with absence of 
lateral walls). Our simulations thus indicate that the lateral confinement does not prevent from the emergence
of the roll regime. Interestingly, with the gap width used here (i.e., $W=68D$), we always get a single pair of
rolls. We could however conjecture that flow configurations with a much larger gap width should give rise to the formation of
several pairs of rolls. In our configuration, the pair of rolls always exhibit the same direction of rotation, leading
to a downward motion of the particles in the center of the cell and a upward motion at the lateral walls. The roll regime
will be therefore labeled as $R^-$ (the minus sign referring to the downward motion of the grain at the cell center).

Supported flows exhibit striking feature with a dense core floating on a dilute basal layer. This regime has been
first uncovered in the work by Brodu et al. \cite{Brodu2015}. The apparition of a dense core is probably linked to the cluster
instability in granular gas \cite{Zanetti1993}. Importantly, the longitudinal rolls are still present in this flow regime and
are not suppressed by the presence of the dense core. They give rise to particle exchange between the dense core and the dilute
surrounding region. At the onset of the supported regime (i.e., $\theta=25^\circ$ for $H=6D$), the core possesses two planes of symmetry, a vertical and
an horizontal one. However, for larger inclination angles, the horizontal symmetry is broken and the core get bended.
As a result, the core starts to rock back and forth.

\begin{figure}[htb]
\begin{center}
\includegraphics[width=0.5\columnwidth]{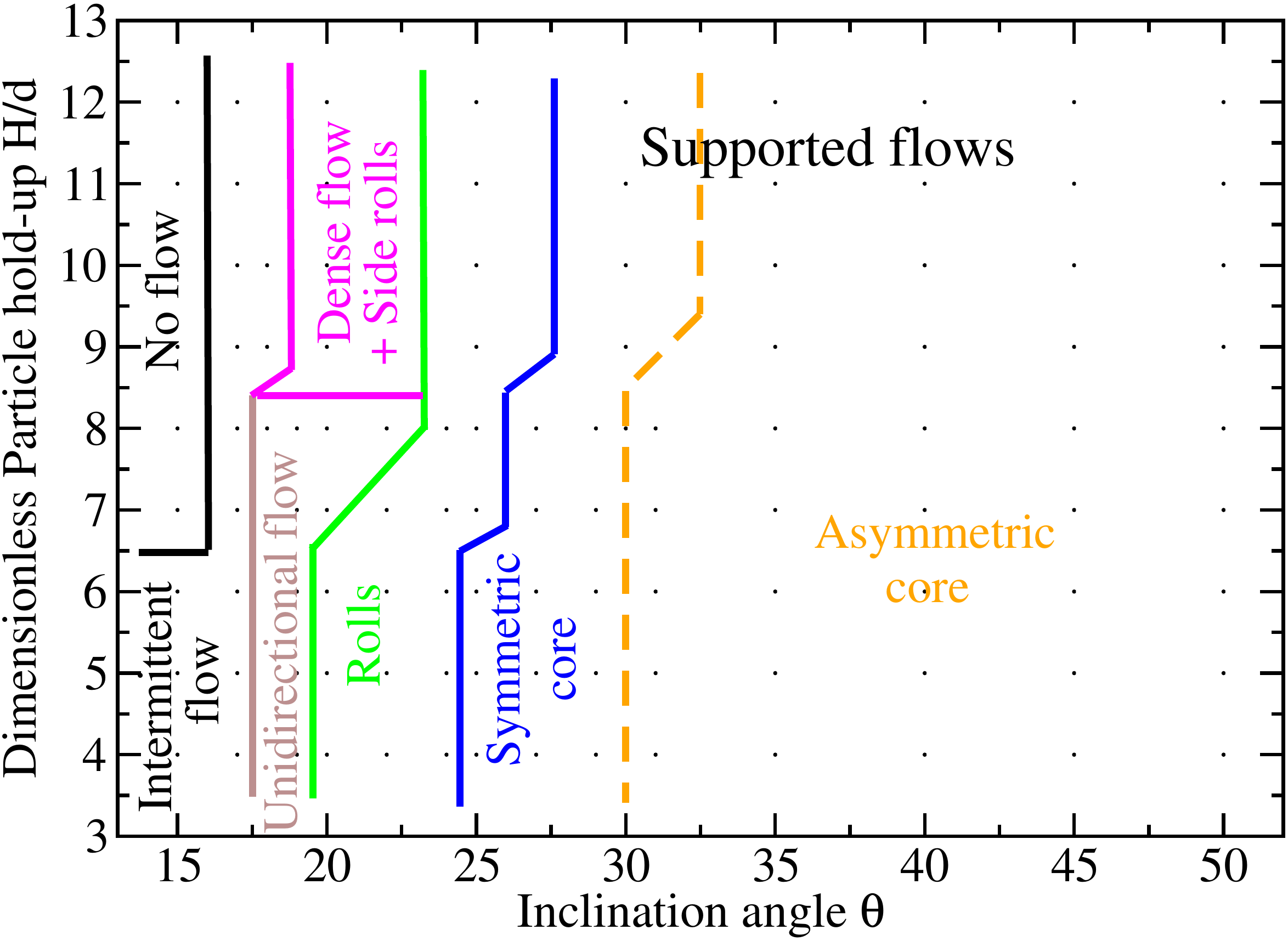}
\end{center}
\caption{Phase diagram in the parameter space $(H,\theta)$ for $W=68D$.  
$U$: unidirectional and dense flows with layering observed close to the base; $SR$: dense flow regime
with small longitudinal vortices located at the flow surface and close to the side walls; 
$R^-$: flows with a pair of longitudinal rolls that spans over the entire cell width; $C$: Supported flows
with a symmetric dense core; $AC$: Supported flows with an asymmetric dense core. }
\label{diag}      
\end{figure}
The above flow regimes are all steady and fully developed and have a limited domain of existence in the parameter space $(H,\theta)$ 
as illustrated in Fig.~\ref{diag}. Several remarks follow. First, at low angles (i.e., $\theta \le 17^\circ$), the flow is not steady: 
the mean flow velocity does not reach a steady value but fluctuates a lot. These flows are close to the jamming transition and 
have been named as intermittent flows.
Second, we can note that the inclination angle is the parameter which drives the transition of the different flow regimes.
As the inclination angle is increased, several transitions occurs successively: at roughly $20^\circ$ unidirectional flows ($U$) give
rise to roll regime ($R^-$) which itself leads to supported flow above $25^\circ$. The critical angles characterizing these transitions
increases slightly with increasing particle hold-up. We will describe carefully these transitions in Section~5.

\section{Packing fraction, velocity and temperature profiles}
\begin{figure}[htb]
\begin{center}
\subfigure[]{\includegraphics[width=0.3\columnwidth]{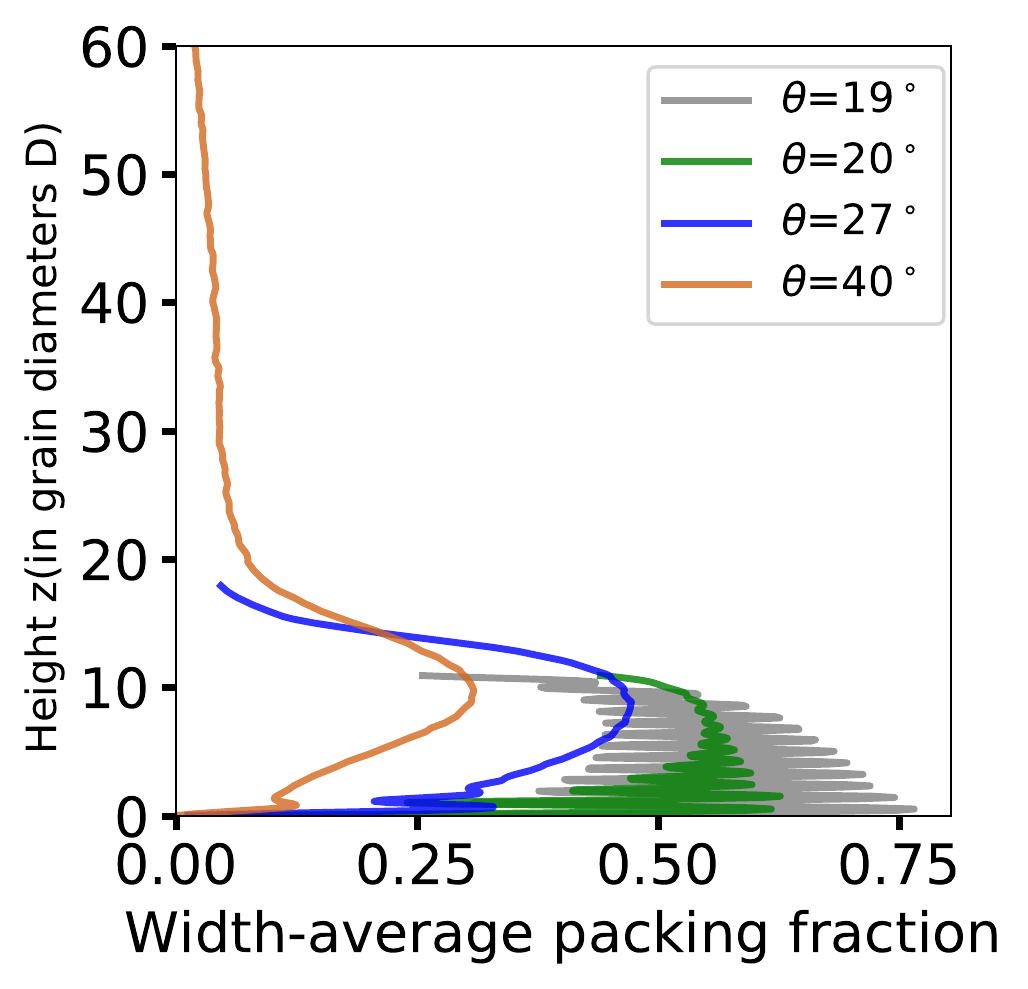}}
\subfigure[]{ \includegraphics[width=0.3\columnwidth]{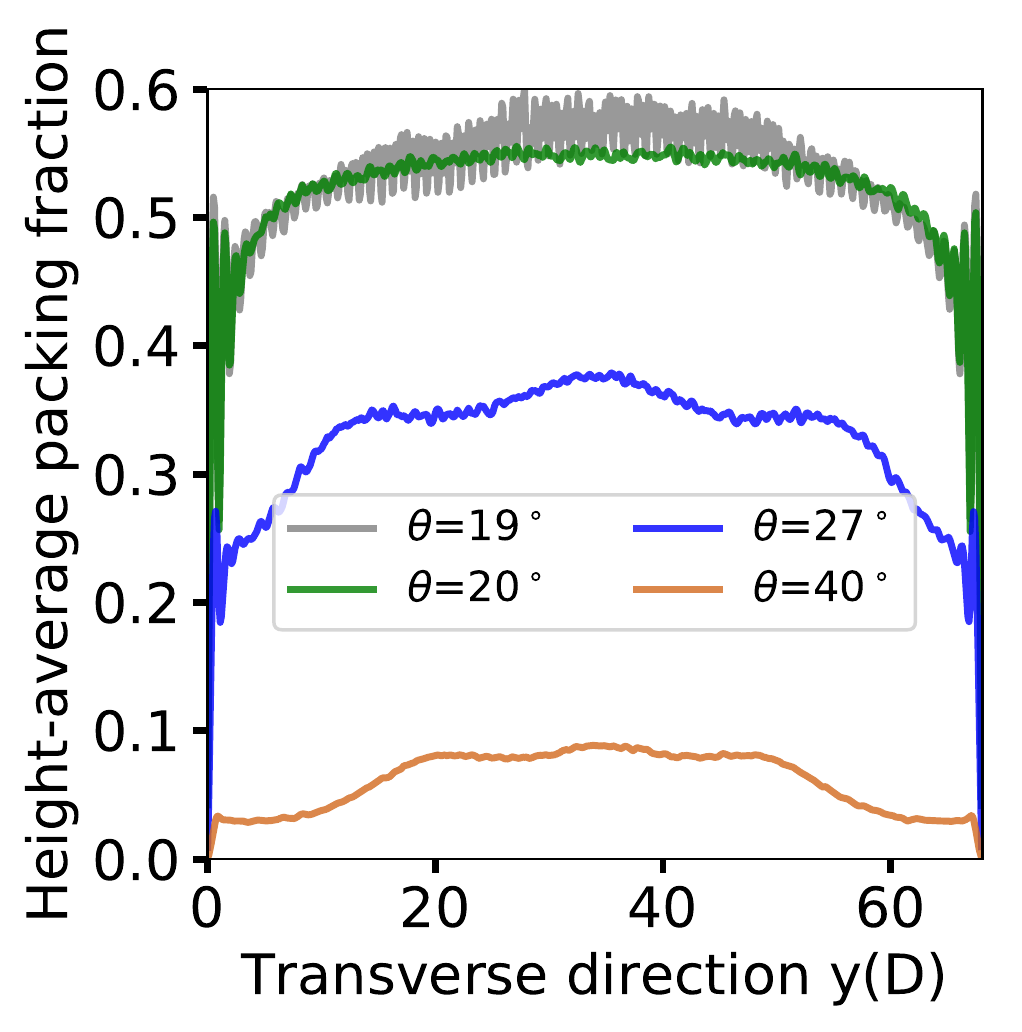}}\\
\subfigure[]{ \includegraphics[width=0.3\columnwidth]{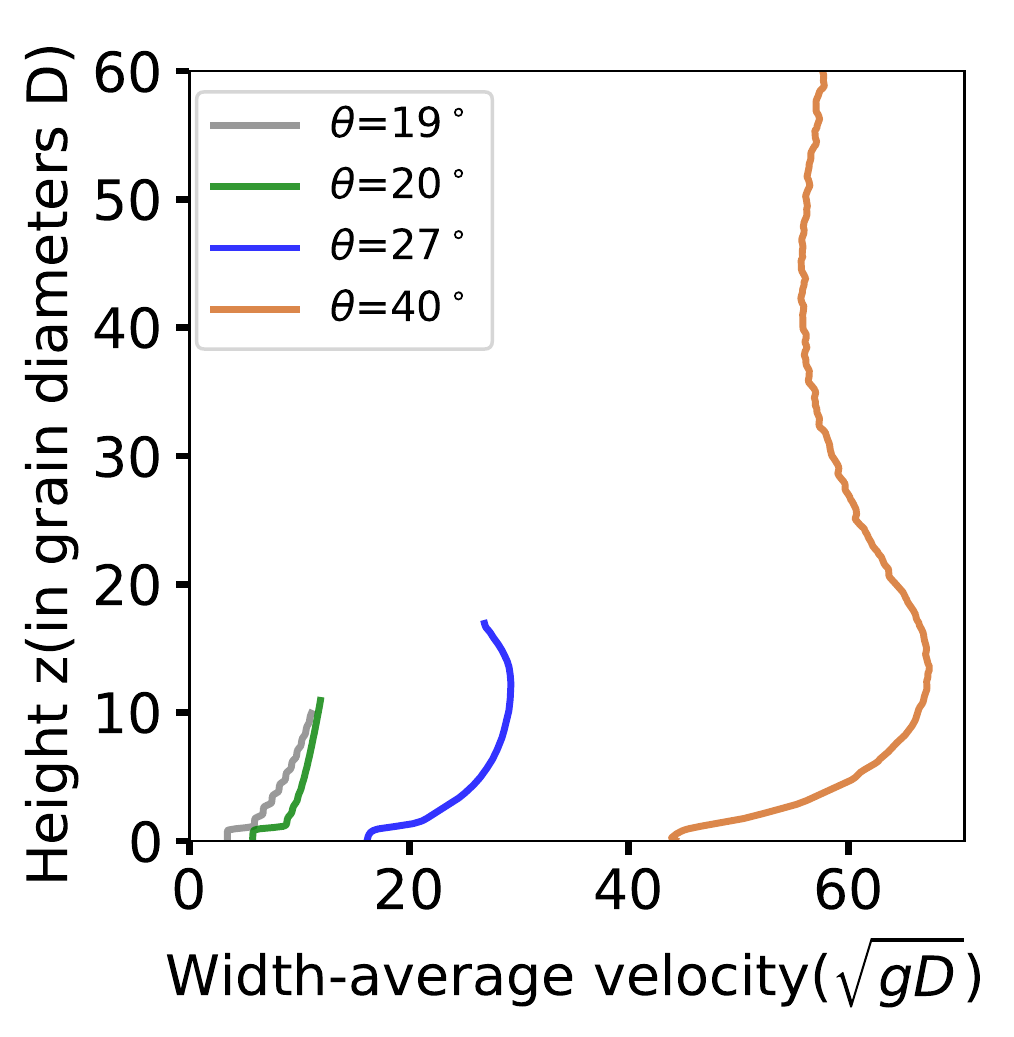}}
\subfigure[]{ \includegraphics[width=0.3\columnwidth]{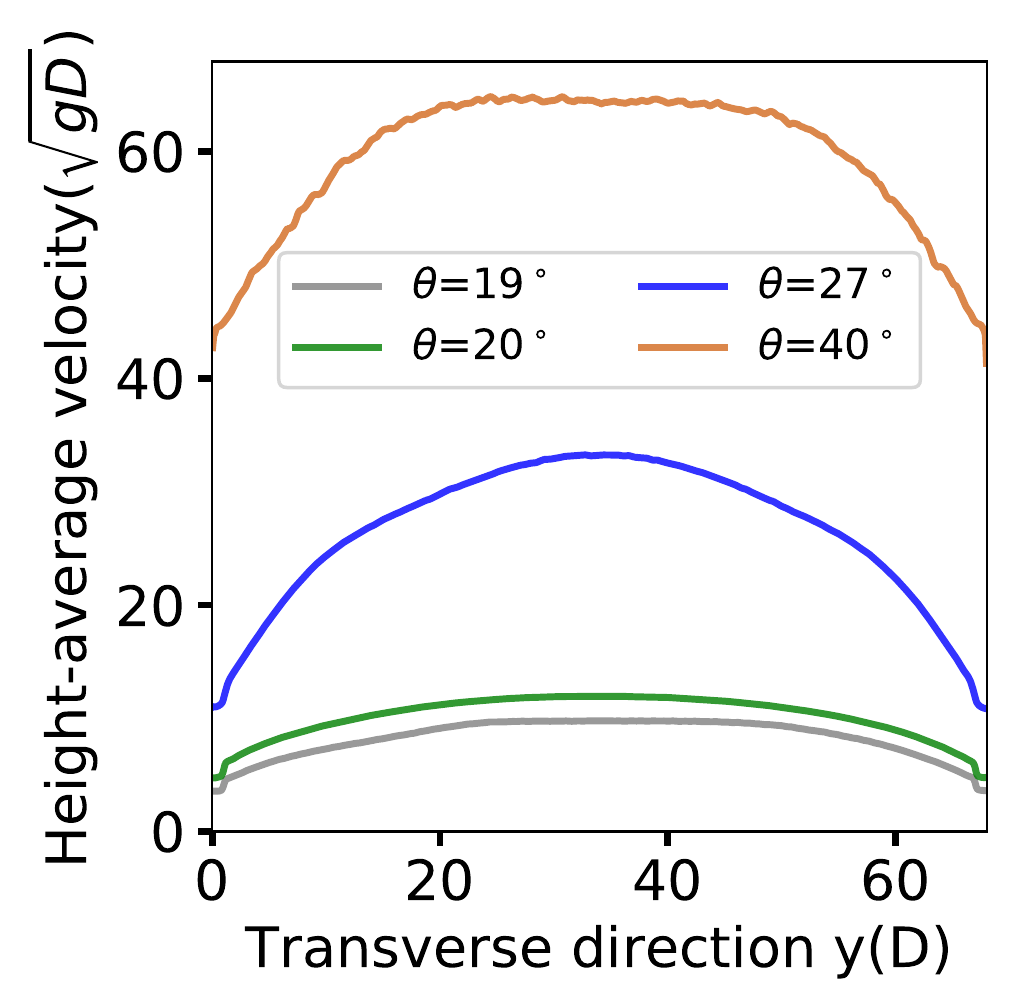}}
\end{center}
\caption{ 
(a) Vertical and (b) transverse profiles of the packing fraction for different inclinations angles for $H=6D$;
(c) Vertical and (d) transverse profiles of the particle stream-wise velocity for $\theta=19,20,27$ and $40^\circ$ and a fixed
particle hold-up $H=6D$.
}
\label{packing_velocity}
\end{figure}

Vertical and transverse packing fraction profiles as well as stream-wise particle velocity profiles for different flow regimes are displayed 
in Fig.~\ref{packing_velocity}. As expected,
the flow velocity increases with increasing angle. We can note however that the increase is not only due to an increase of
the shear rate but also to a large augmentation of the slip velocity at the boundaries. In the vertical direction,
the flow is sheared over the whole flow depth at low inclination angles
(i.e., $19^\circ$ and $20^\circ$), while
the shear zone is essentially localized in the dilute layer close to the bottom at higher angles (i.e., for supported flows),
In the transverse direction, similar features are observed. At low inclination angles, the flow is sheared almost uniformly over the whole width.
In contrast, at larger angles, the shear rate is more pronounced in the dilute layer close to the vertical walls than within
the dense core. At $\theta=40^\circ$, the dense core flows as a plug and does not exhibit any shear within it. 

\begin{figure}[htb]
\begin{center}
\subfigure[]{ \includegraphics[width=0.3\columnwidth]{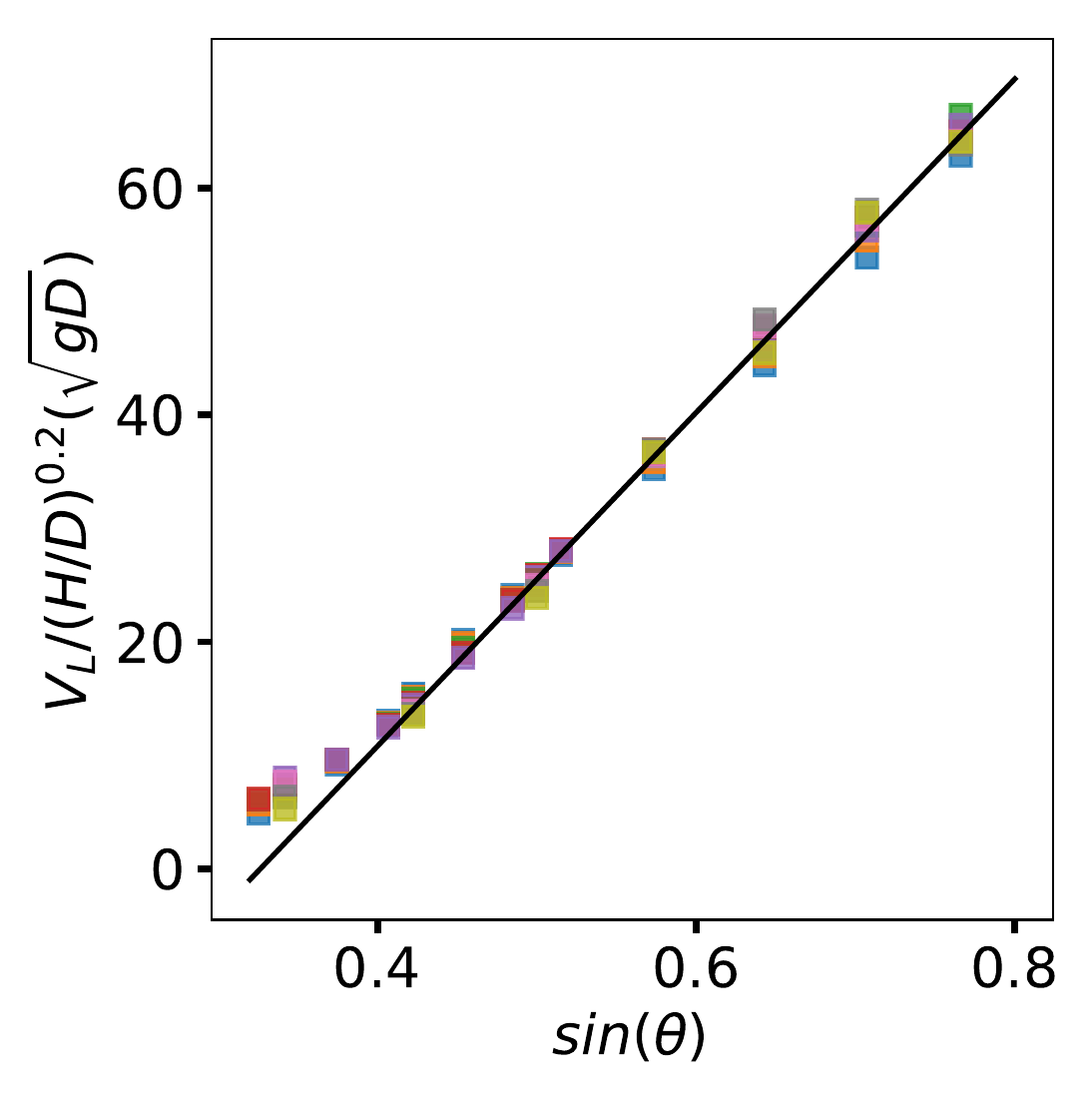}}
\subfigure[]{ \includegraphics[width=0.3\columnwidth]{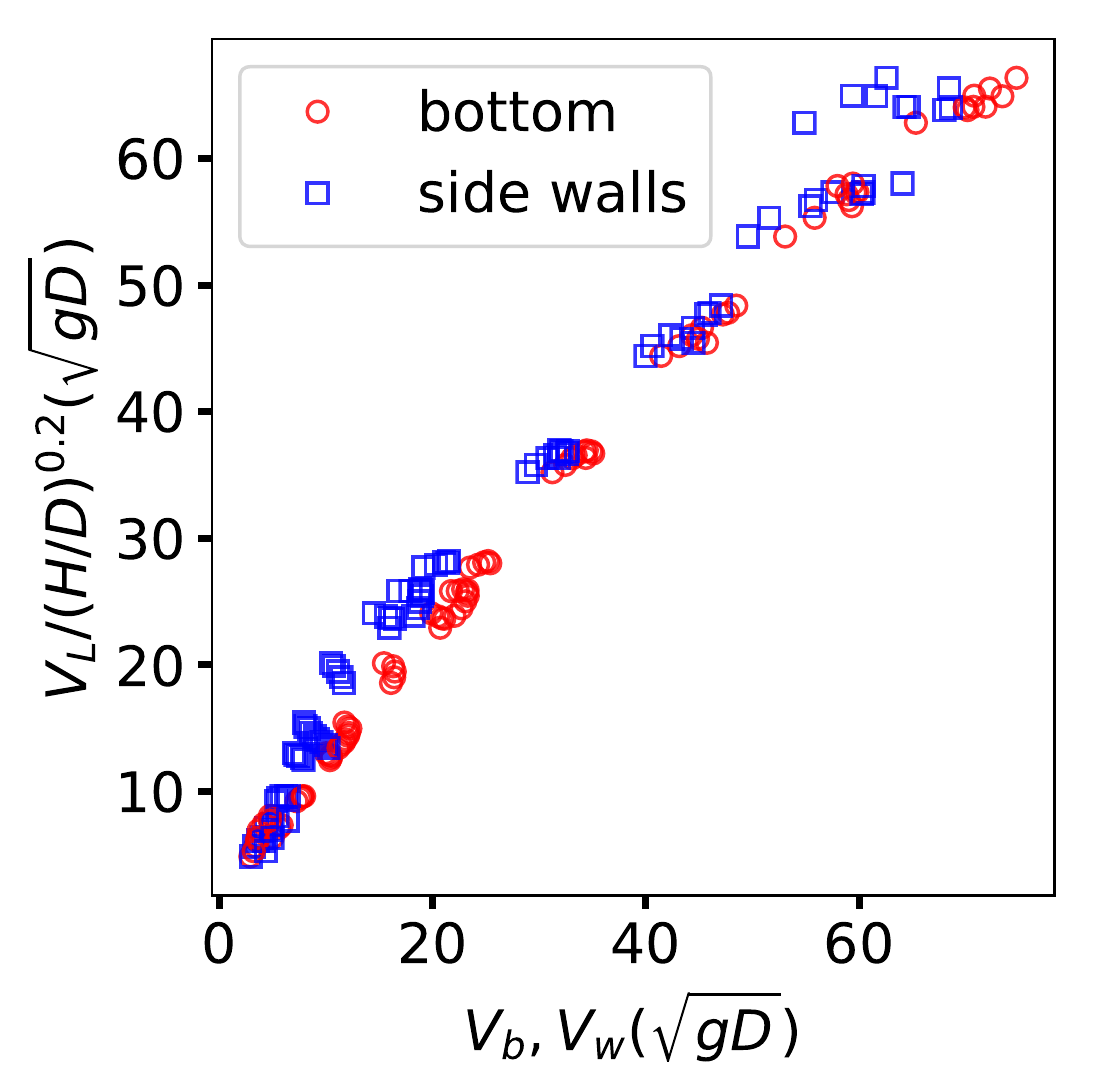}}
\end{center}
\caption{ 
(a) Rescaled mean flow velocity $V_L/(H/D)^{0.2}$ as a function the inclination angle. 
(b) Bottom (circle symbols) and (square symbols) side-wall velocities versus the rescaled mean flow velocity $V_L/(H/D)^{0.2}$.
}
\label{scaling_vel}
\end{figure}

Importantly, we confirm the scaling law proposed by \cite{Brodu2015}
concerning the mean flow velocity $V_L$:
\begin{equation}
\frac{V_L}{\sqrt{gD}} \approx (H/D)^{\alpha} \left( A + B \sin \theta \right)
\end{equation}
with $\alpha \approx 0.2$, $A\approx 147$ and $B\approx -48$. The value of the exponent $\alpha$ reported in \cite{Brodu2015} was
a bit larger ($\alpha \approx 0.25$). Here we find a better collapse of the data with $\alpha\approx 0.2$ (see Fig.~\ref{scaling_vel}.a).
This scaling indicates that the mean velocity increases both with the inclination angle and the particle hold-up. However, it is important
to note that the increase of the mean flow velocity with the particle hold-up is rather mild and drastically differs from the Bagnold scaling
law (i.e.,$V_L \propto H^{3/2}$) which is  relevant for unconfined dense granular flows. 

\begin{equation}
V_b \approx V_w \approx V_L/(H/D)^{0.2} \approx \left( A + B \sin \theta \right)\sqrt{gD}
\label{vbw_eq}
\end{equation}
In addition to the mean flow velocity, the velocities at the boundaries are also interesting and relevant quantities. We find that
the velocity at the bottom and at the side walls are almost independent on the particle hold-up within the range investigated so far
(i.e., $4<H/D<12$) but increases with increasing inclination angle. Interestingly, they are quantitatively similar and
are linearly correlated with the rescaled flow velocity $V_L/(H/D)^{0.2}$ as illustrated in Fig.~\ref{scaling_vel}b:

\begin{figure}[htb]
\begin{center}
\subfigure[]{\includegraphics[width=0.3\columnwidth]{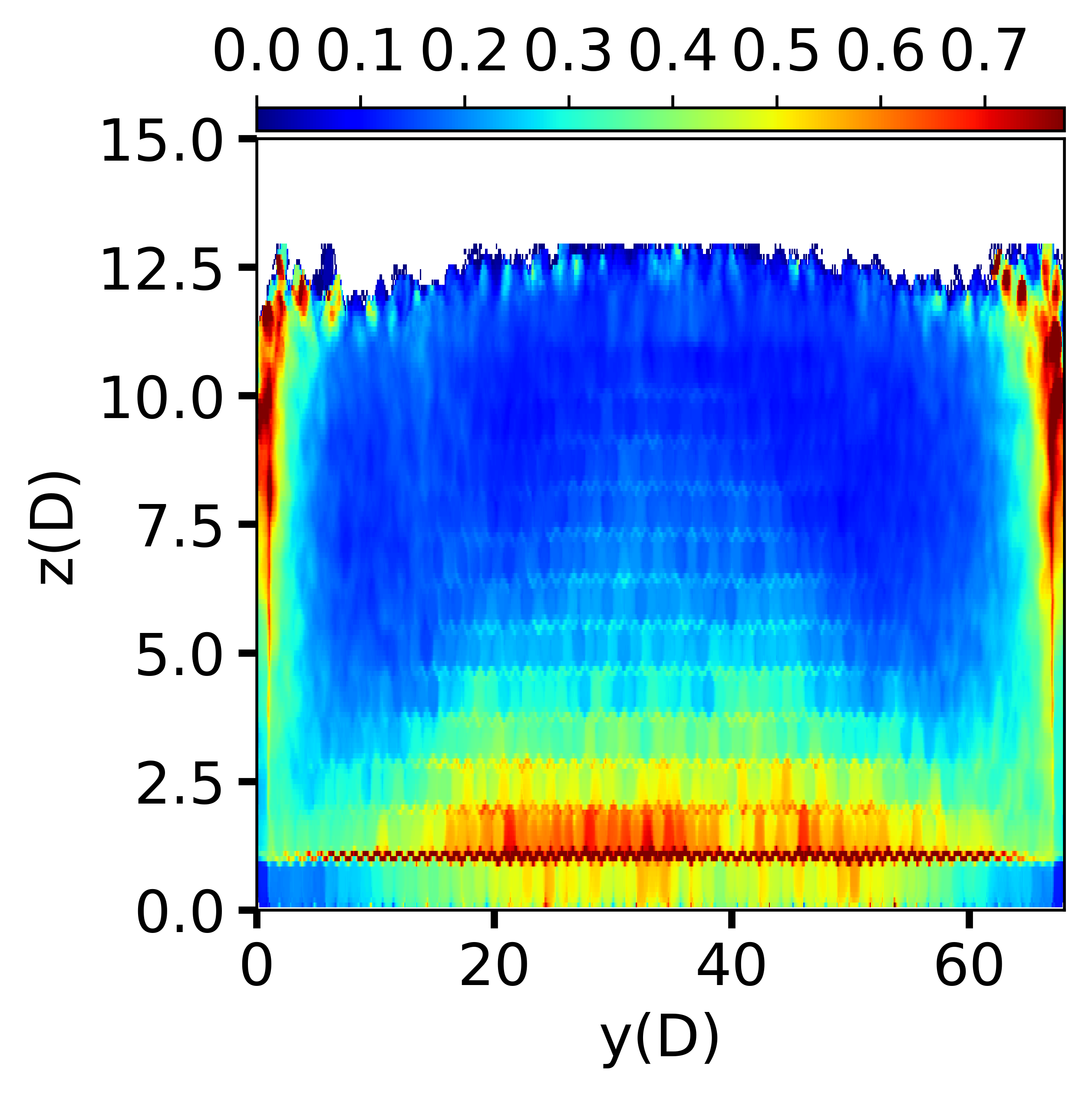}}
\subfigure[]{\includegraphics[width=0.3\columnwidth]{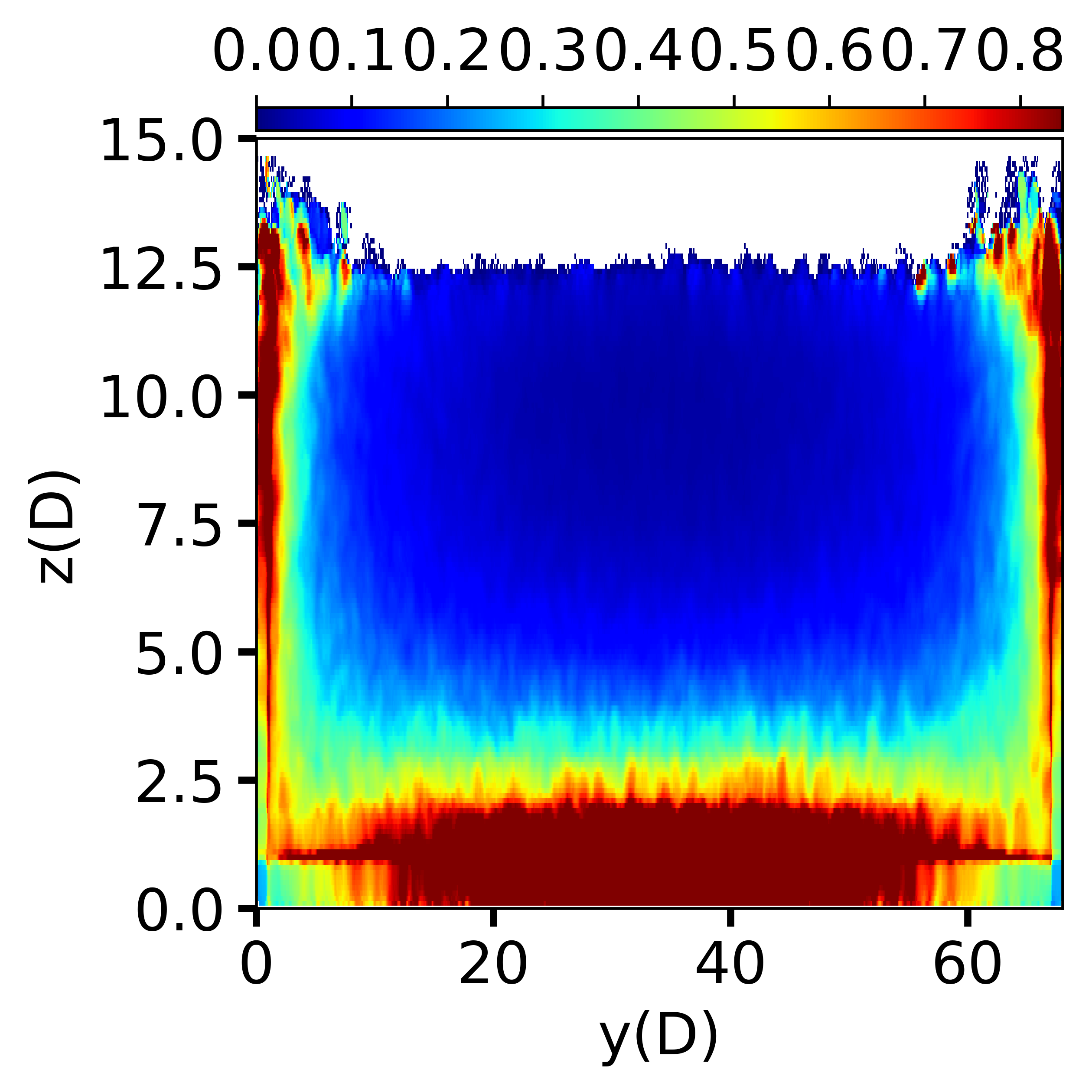}}
\subfigure[]{\includegraphics[width=0.3\columnwidth]{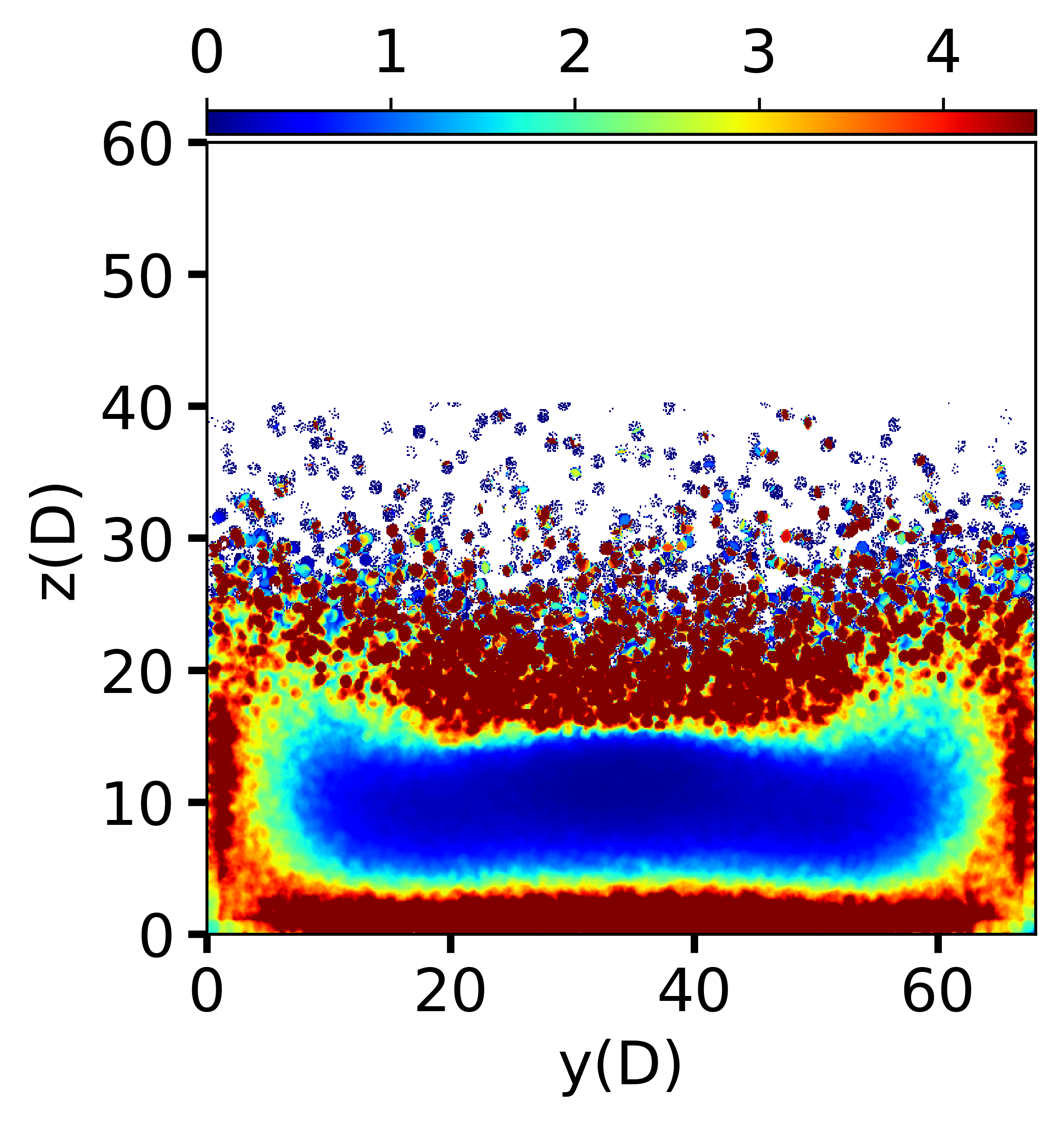}}
\subfigure[]{ \includegraphics[width=0.3\columnwidth]{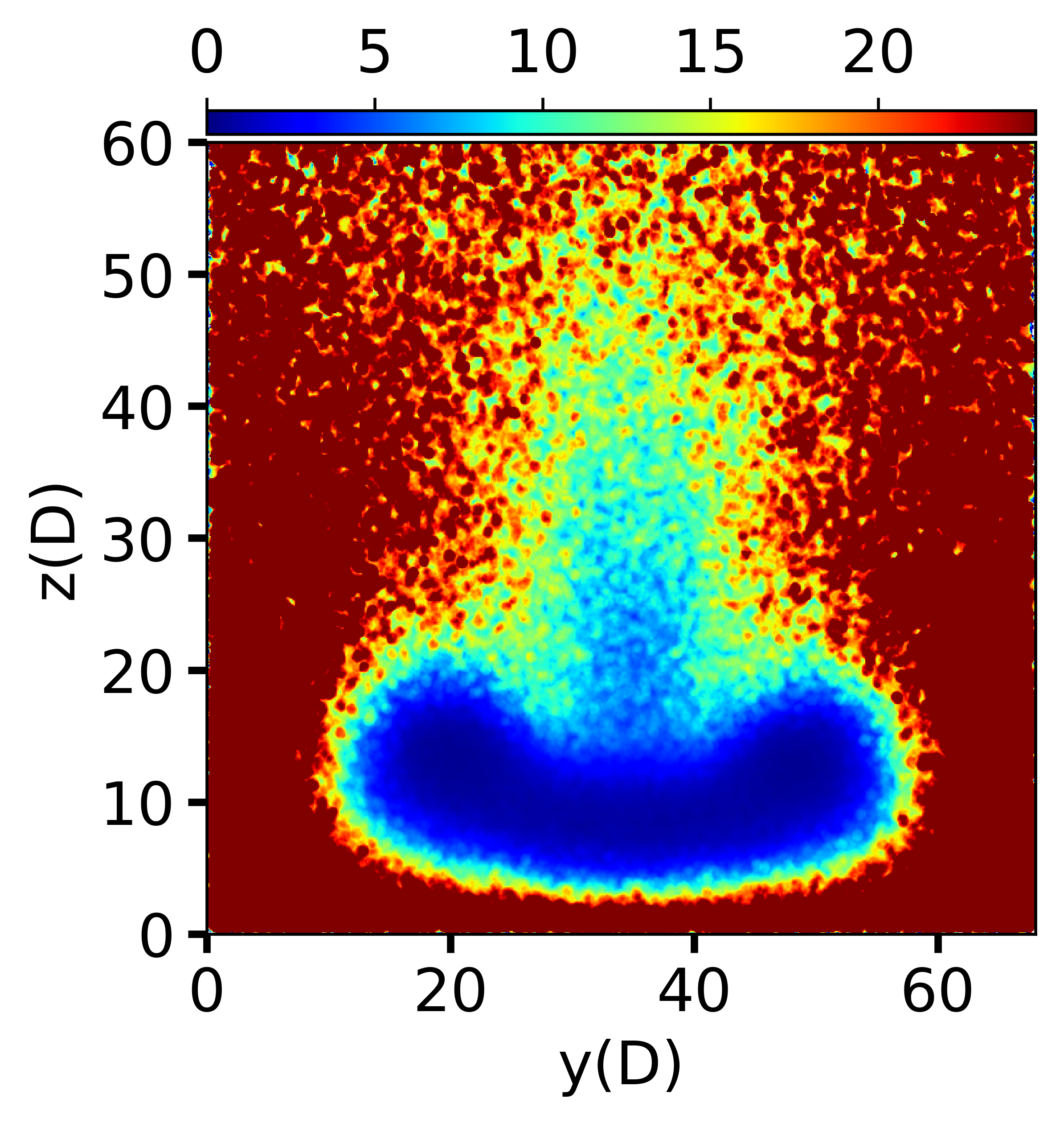}}
\subfigure[]{\includegraphics[width=0.3\columnwidth]{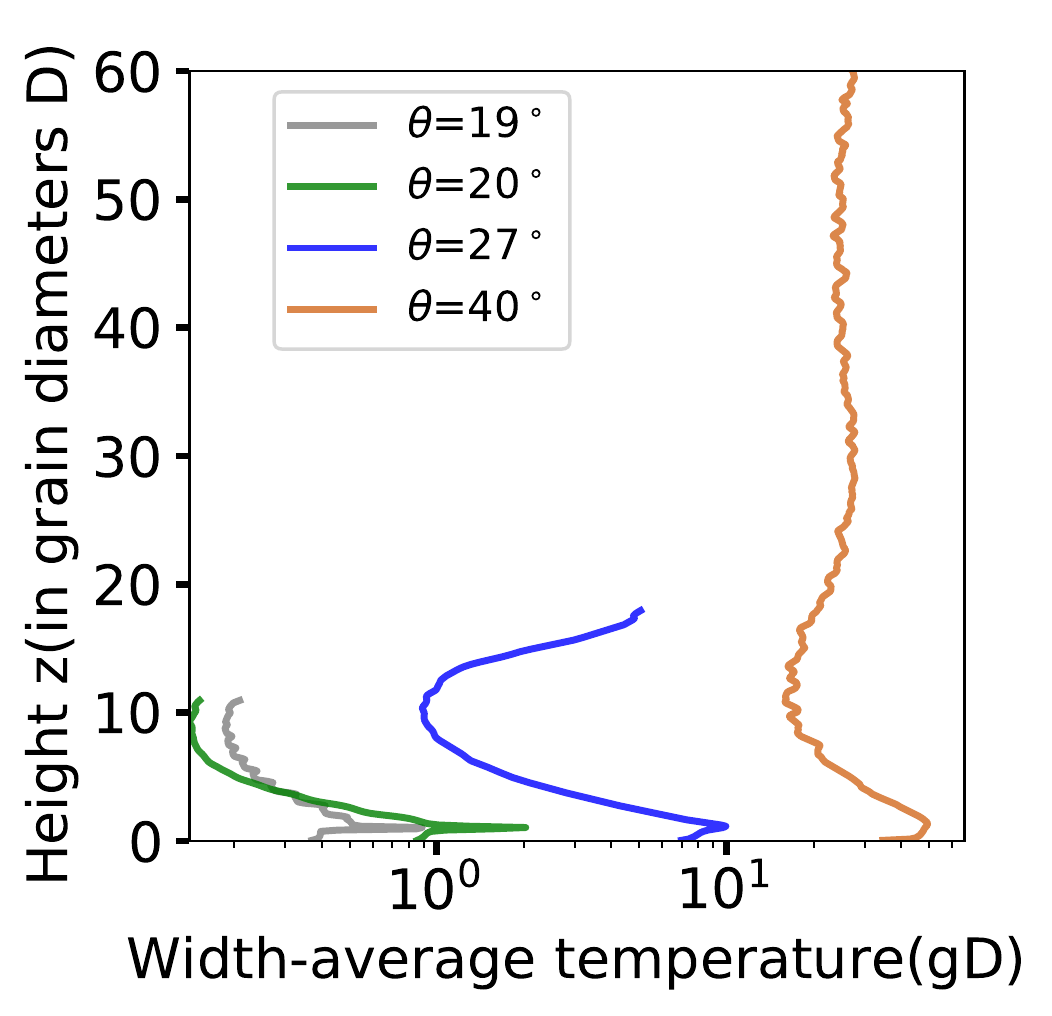}}
\subfigure[]{\includegraphics[width=0.3\columnwidth]{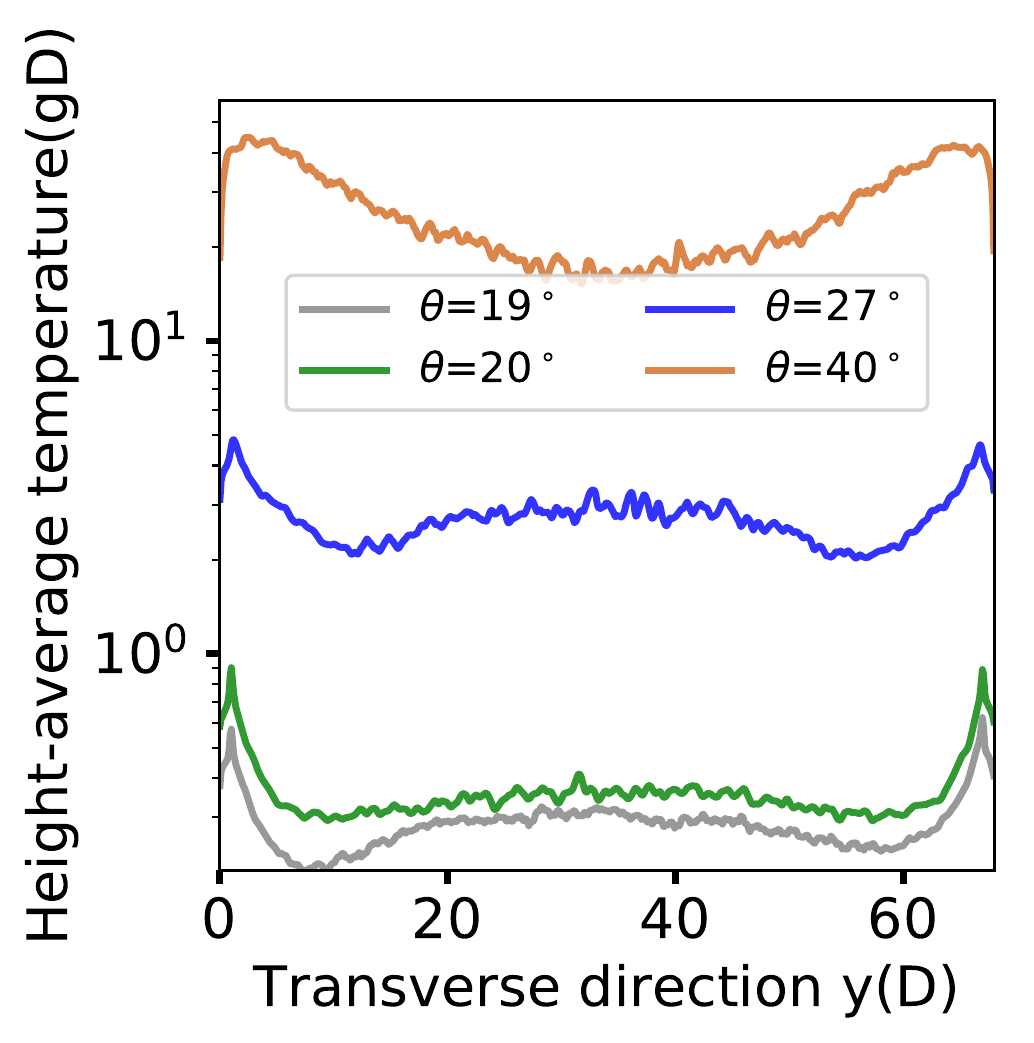}}
 \end{center}
\caption{ Temperature map for different flow regimes. $H=6D$ and $W=68D$.
(a) Unidirectional dense regime;
($\theta=19^\circ$); (b) Roll regime ($\theta=20^\circ$): (c) and (d) supported flows: symmetric core ($\theta=27^\circ$) and
asymmetric core ($\theta=40^\circ$); 
(e) Corresponding vertical and (f) transverse profiles of the granular temperature.}
\label{temp}
\end{figure}
Granular temperature is a measure of the particle velocity fluctuations. It is
an important parameter in various theories aiming to capture granular flow behaviors.
It is defined as $T = (T_{xx} + T_{yy} + T_{zz}) /3$
where $T_{ij} = <u_i u_j> - <u_i><u_j>$, $u_i$ is the $i$ component of the instantaneous particle velocity  and $<..>$ stands for time averaging 
and spatial averaging in the stream-wise direction.
We provide in Fig.~\ref{temp} temperature map within the cross-section of the flow as well as vertical and transverse
profile of the temperature for various flow regimes.
We observe contrasting features for slow and large angles, respectively. For unidirectional flows, the temperature is relatively homogeneous with
a temperature at the bottom slightly greater than within the bulk flow (see Fig.~\ref{temp}.a).
In the roll regime, the temperature is still very homogeneous within the bulk flow
but there is a larger contrast of temperature between the bottom temperature and the bulk one.
For large angles (i.e., for supported flows), the temperature map exhibits contrasting features.
The supported dense core is very cold and surrounded by a dilute hot gas. These flow regime thus display strong heterogeneities
of temperature which is strongly correlated to particle volume fraction.

\section{Flow regime transition}
In this section, we describe carefully the transition between the different flow regimes. For that purpose,
we investigate the variation of several key parameters when we vary the inclination angle. 

\begin{figure}[htb]
\begin{center}
\includegraphics[width=0.4\columnwidth]{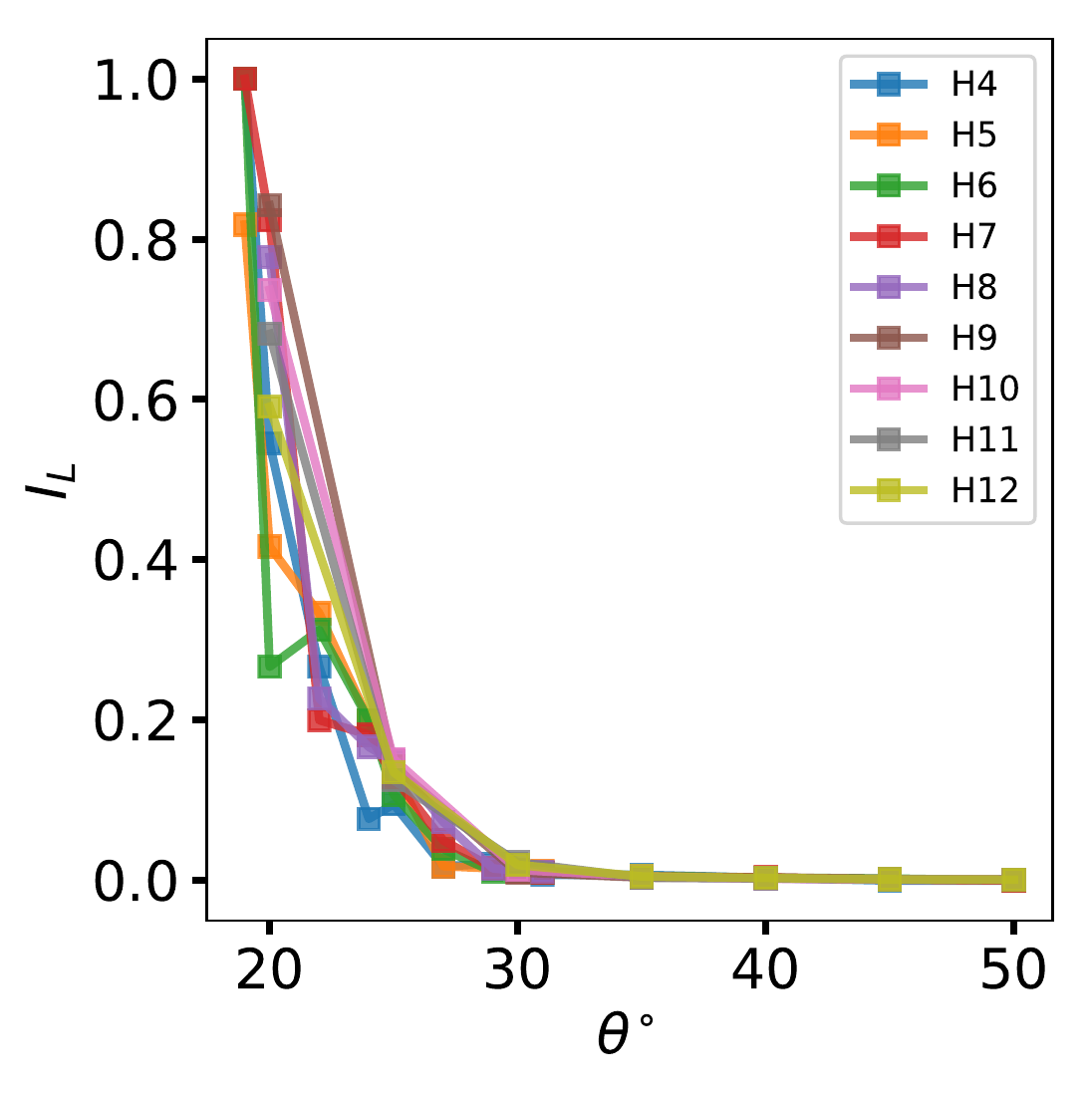}
\end{center}
\caption{
Layering index $I_L$ as a function of the inclination angles for various particle hold-up: 
$I_L=(1/n)\sum_i Y\left( <\phi>_i-\phi_i^{min}-0.05 \right)$ where $Y$ is the Heaviside function.
}
\label{layer_index}
\end{figure}
We first focus on a parameter characterizing the layering process seen mainly in the undirectional and dense flow regime. 
We introduce a layering index defined as $I_L=(1/n)\sum_i Y\left( <\phi>_i-\phi_i^{min}-0.05 \right)$, where $n$ is the number of layers
of one grain height within the flow, $<\phi>_i$ the averaged packing fraction over the layer $i$, $\phi_i^{min}$ is the minimum
value of the packing fraction $\phi(z)$ within the layer $i$, $Y$ is the Heaviside function and $0.05$ is a threshold value
for quantifying the layering.  The variation of the layering index with the inclination 
angle 
is shown in Fig.~\ref{layer_index}. SFD unidirectional and dense flows exhibit a strong layering with a layering index close to $1$.
Upon increasing the inclination angle (from $20^\circ$ to $25^\circ$), the layering index decreases progressively to zero.
Above $25^\circ$ (i.e., in the supported flow regime), the layering has completely disappeared. This index can thus be employed
to characterize the transition towards supported flows. We will see however below that the packing fraction is also a relevant
parameter to identify the supported regime transition.

\begin{figure}[htb]
\begin{center}
\subfigure[]{\includegraphics[width=0.3\columnwidth]{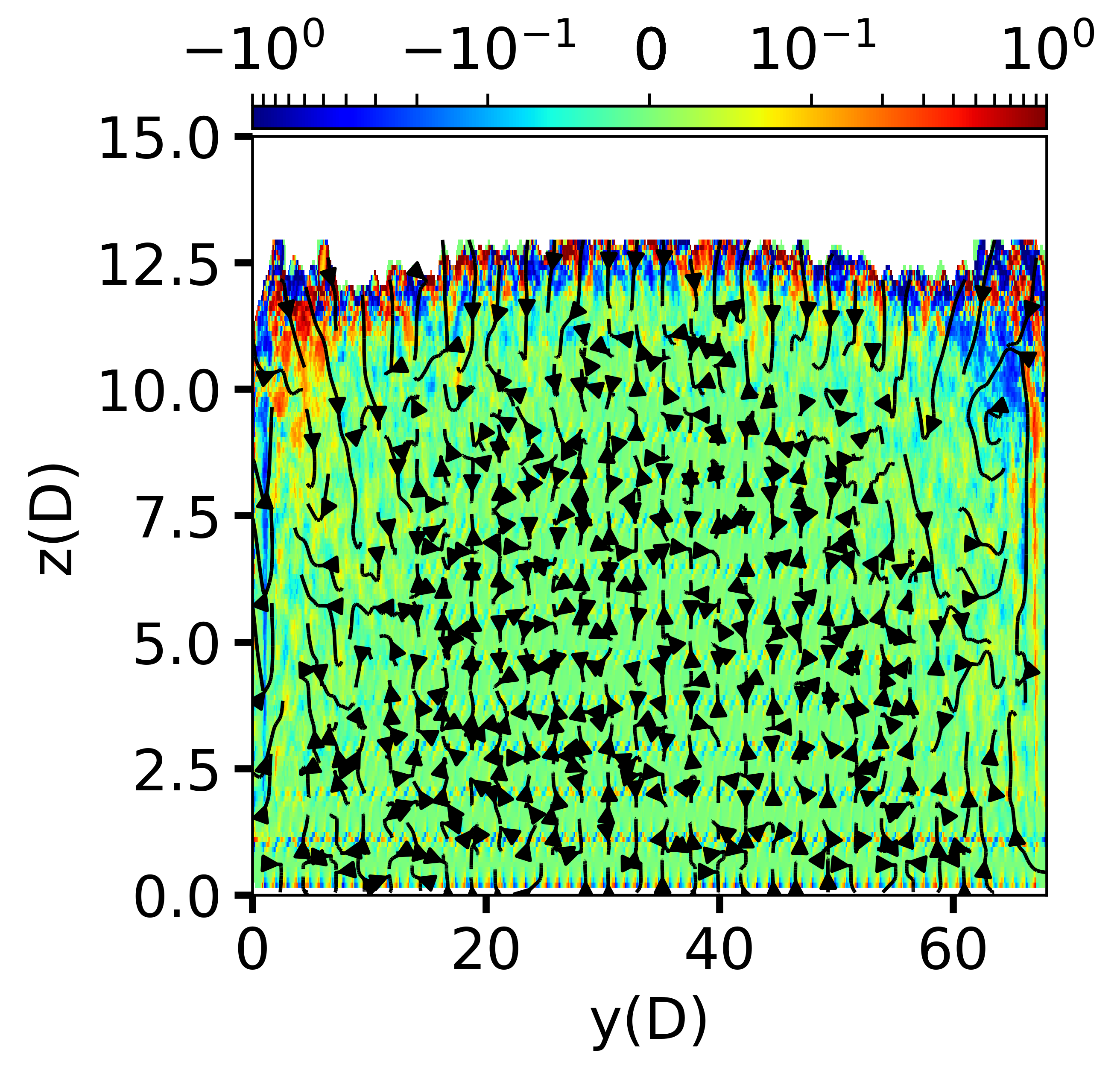}}
\subfigure[]{\includegraphics[width=0.3\columnwidth]{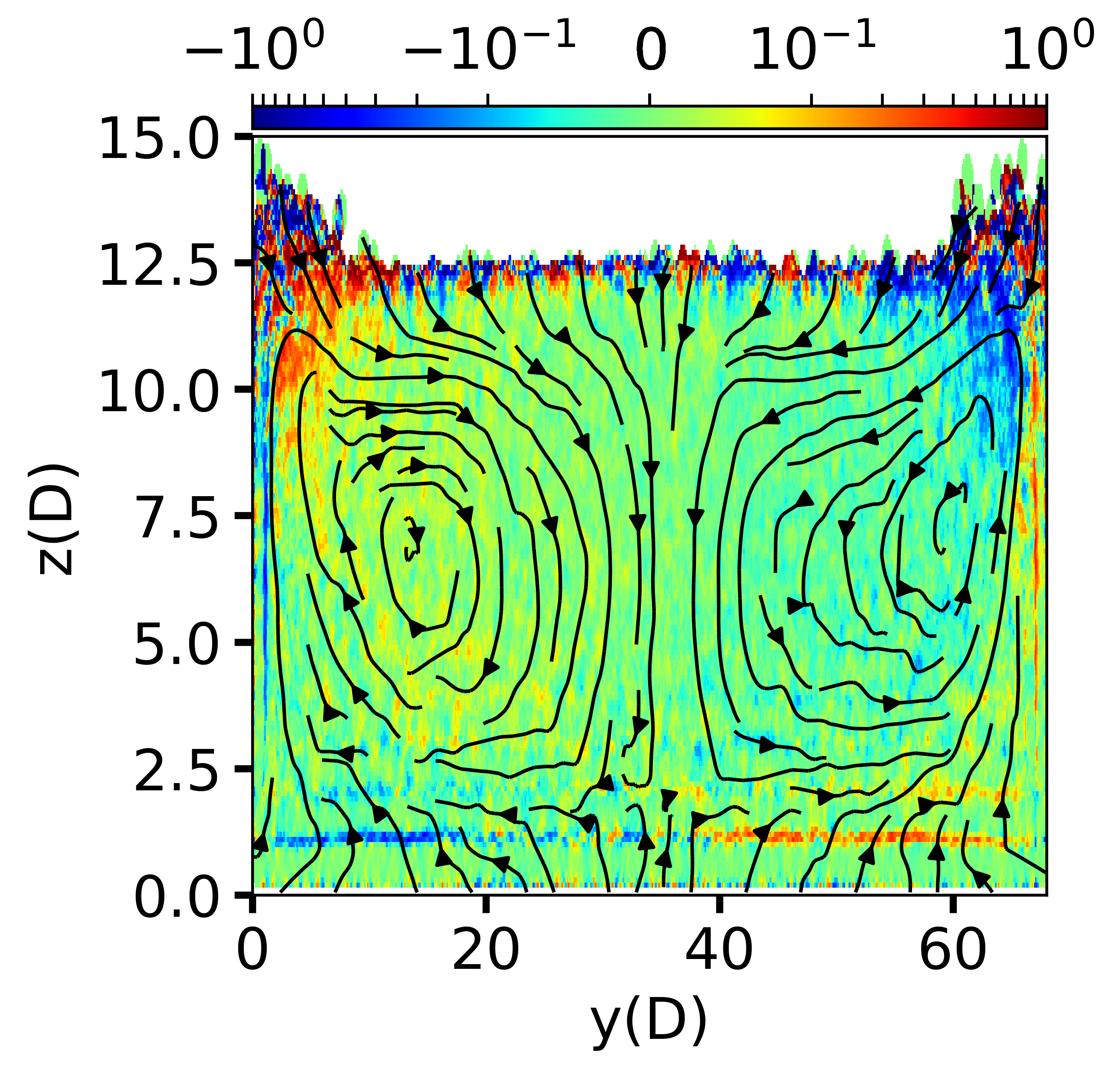}}\\
\subfigure[]{\includegraphics[width=0.3\columnwidth]{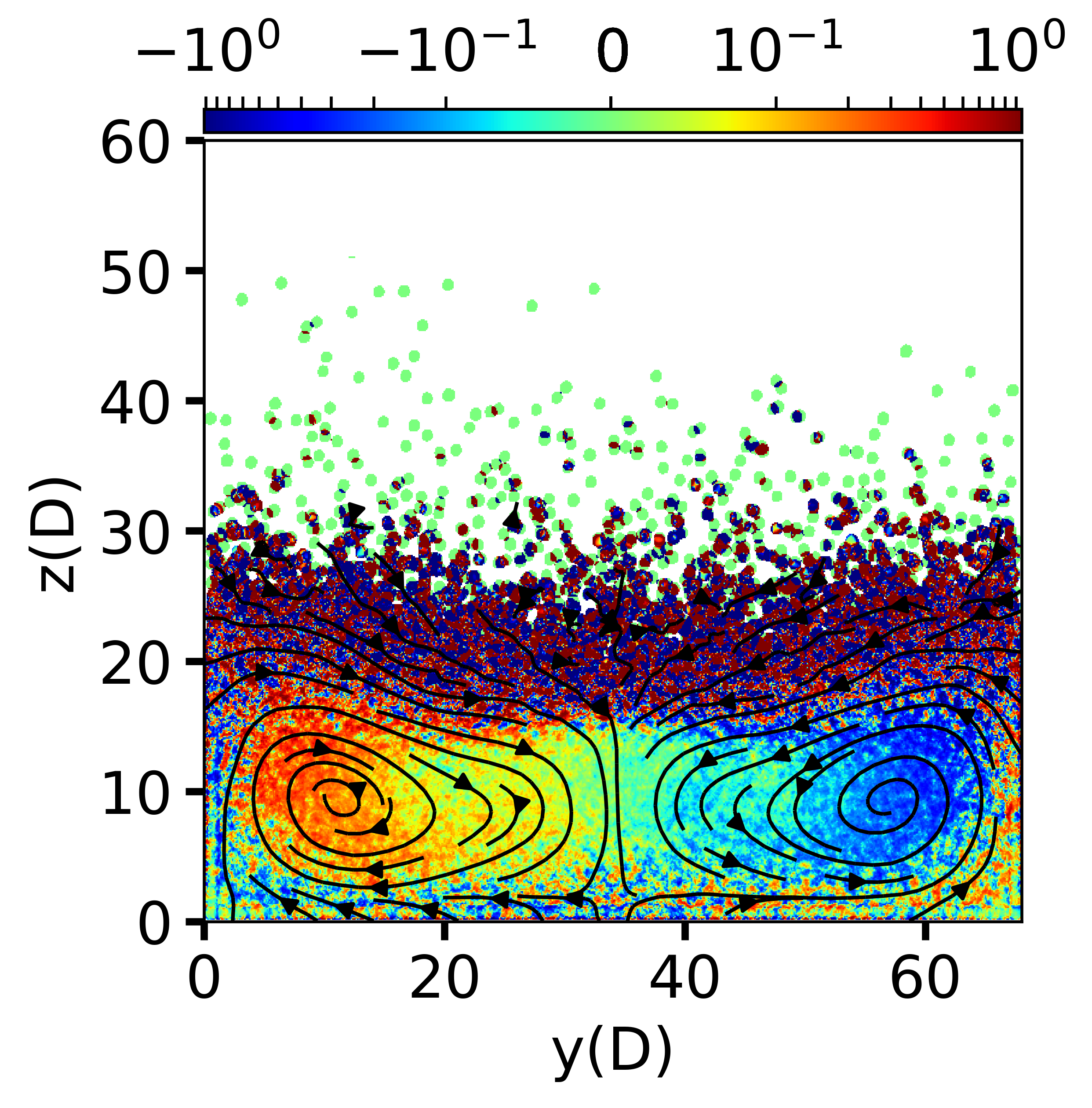}}
\subfigure[]{\includegraphics[width=0.3\columnwidth]{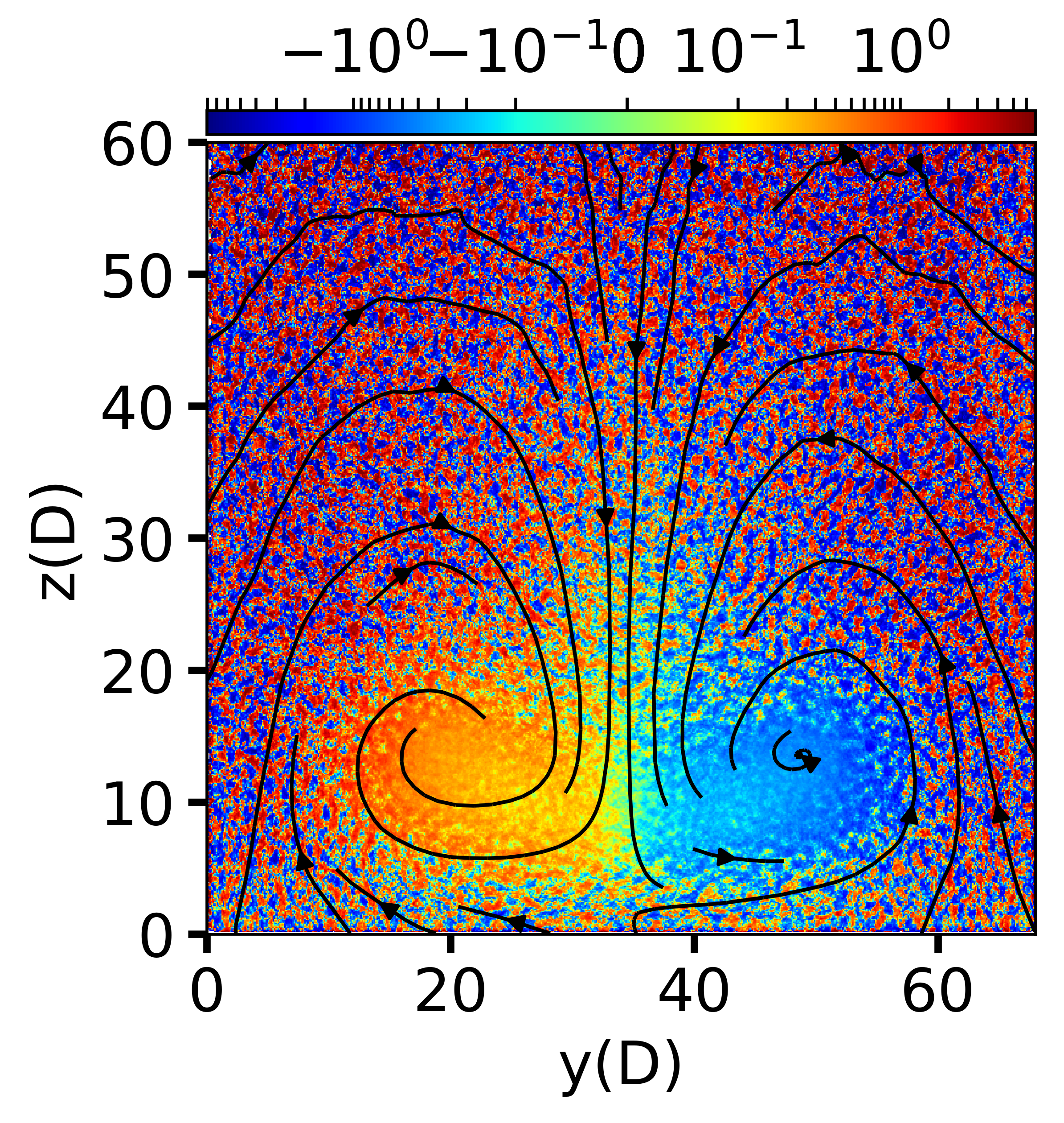}}
\end{center}
\caption{ Vorticity map for different flow regimes. $H=6D$ and $W=68D$. Solid lines represents the streamline in the flow
cross-section. (a) Unidirectional dense regime;
($\theta=19^\circ$); (b) Roll regime ($\theta=20^\circ$): (c) and (d) supported flows: symmetric core ($\theta=27^\circ$) and
asymmetric core ($\theta=40^\circ$). }
\label{vort_map}
\end{figure}
\begin{figure}[htb]
\begin{center}
\subfigure[]{\includegraphics[width=0.35\columnwidth]{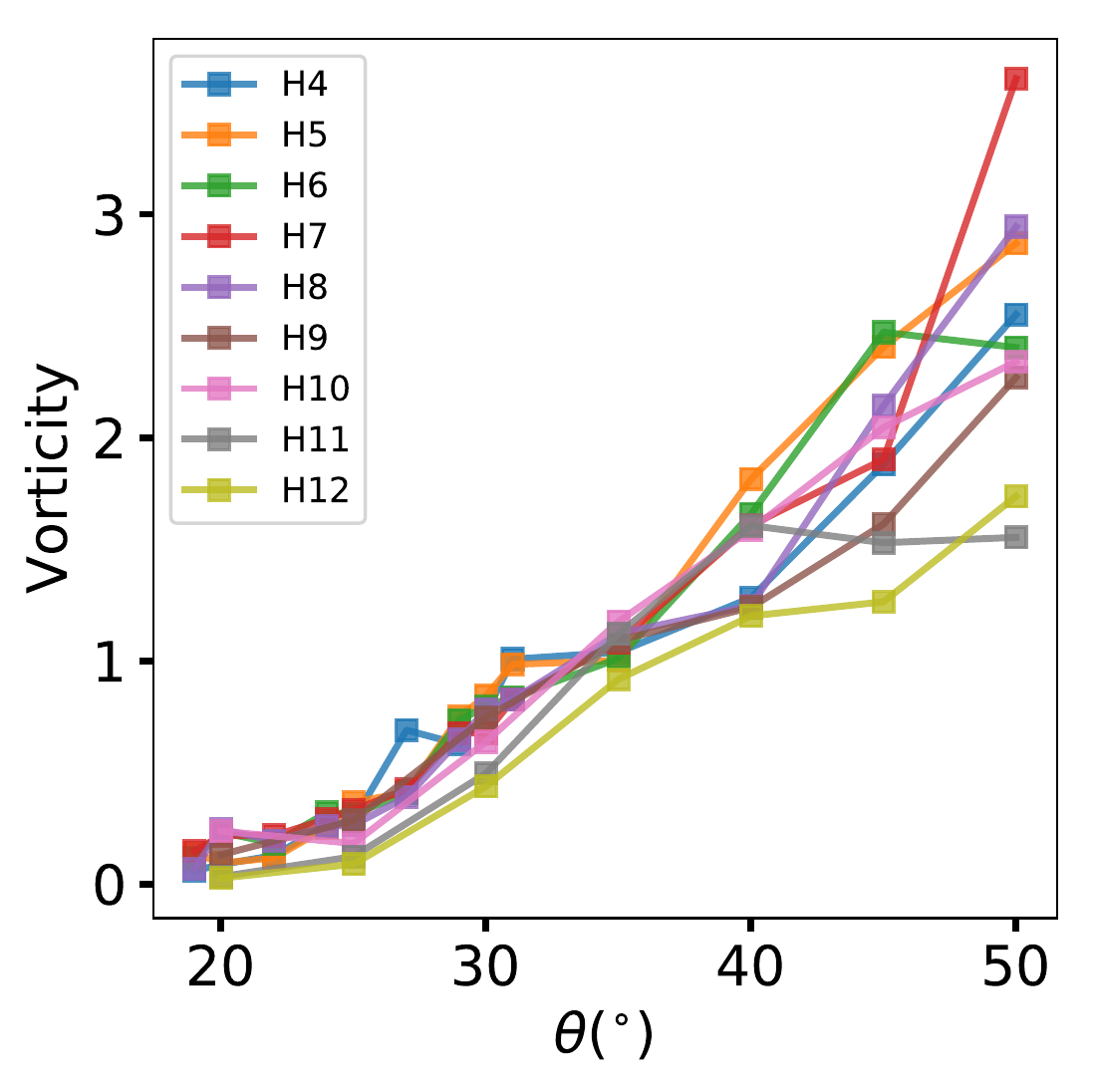}}
\caption{Vorticity versus inclination angle for various particle hold-up. The value of the vorticity stands the maximum value
of the vorticity within the rolls (see Fig.~\ref{vort_map}). }
\label{vorticity}
\end{center}
\end{figure}

The vorticity defined as $\vec{\Omega}=\nabla \times \vec{v}$ is an interesting quantity for characterizing the presence of longitudinal
vortices within the flow. In Fig.~\ref{vort_map}, we present the vorticity map for different flow regimes. For unidirectional flows,
(e.g., $H=6D$ and $\theta=19^\circ$) the vorticity is close to zero (i.e., less than $10^{-1} \sqrt{g/D}$). Upon increasing
inclination angle, roll regime develops with a visible pair of counter-rotative longitudinal vortices (see Fig.~\ref{vort_map}.b) 
but the vorticity does not increases significantly: the maximum in each vortex does not exceed $0.2 \sqrt{g/D}$. Upon further increase of the 
inclination, the flows exhibit similar vorticity pattern
but with increasing values of the vorticity. The behavior is illustrated in Fig.~\ref{vorticity} that presents the maximum value of the
vorticity within the vortex as a function of the inclination angle for various particle hold-up. This plot confirms that
below $20^\circ$ the maximum vorticity is extremely small and progressively increases with increasing inclination angle. 
The vorticity becomes really significant above $25^\circ$ in the regime of supported flows.
As a summary, the vorticity does not provide us with a good parameter for delineating precisely the transition from unidirectional flows
to the roll regime. The vorticity exhibits a too smooth variation across the transition. 

\begin{figure}[hbt]
\begin{center}
\subfigure[]{\includegraphics[width=0.35\columnwidth]{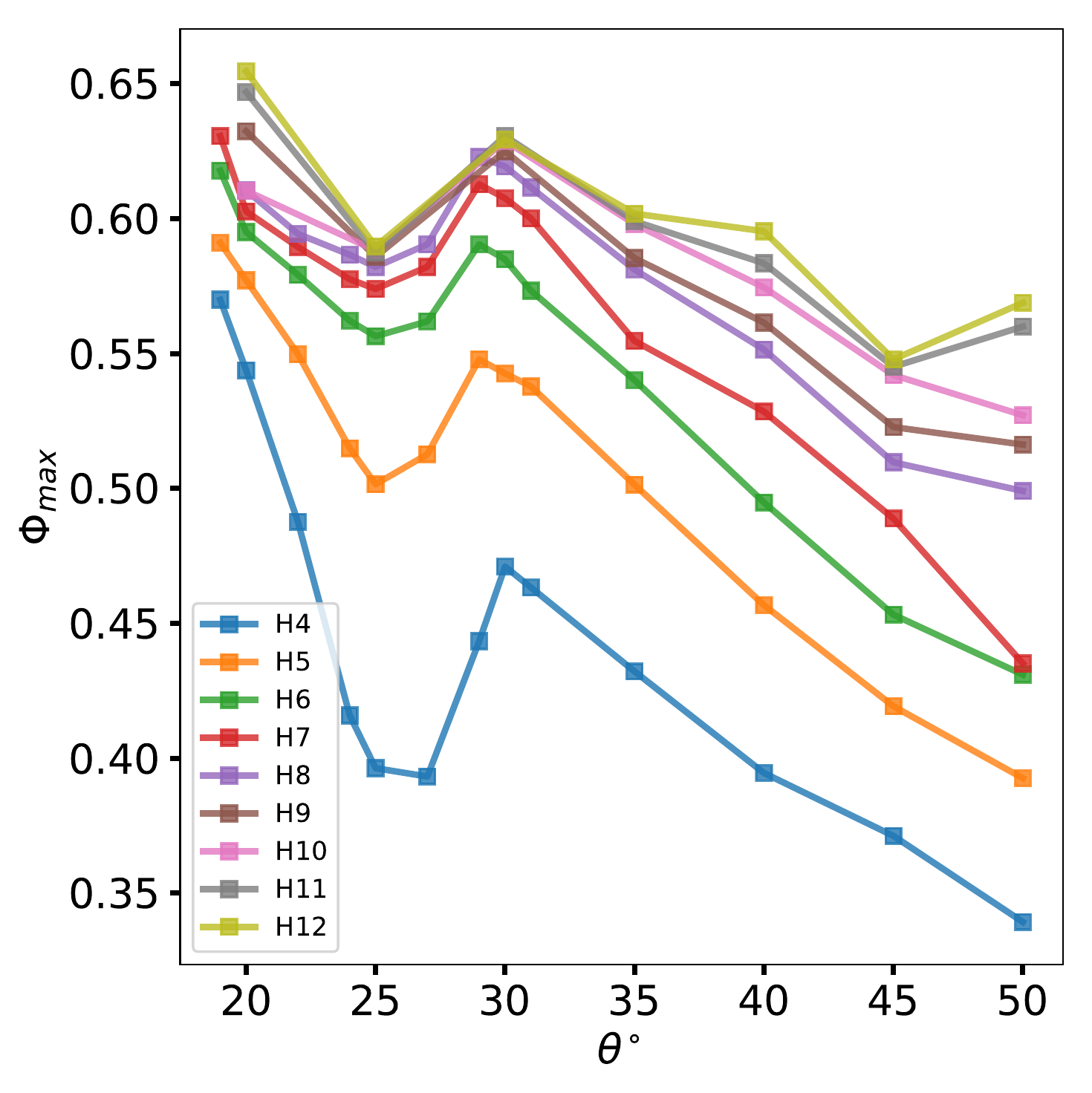}}
\end{center}
\caption{Maximum packing fraction as a function the inclination angle for various particle hold-up.
}
\label{packingmax}
\end{figure}
In addition the layering index, an other parameter can be used to describe the transition from the roll regime to the supported regime. 
As the supported regime is accompanied with the formation of a dense core, it is then natural to investigate how the volume fraction evolves 
with increasing inclination angle.
In Fig.~\ref{packingmax}, we present the maximum value of the volume fraction in the cross-section of the flow as a function of the inclination
angle. For a given particle hold-up, this value first decreases with increasing angle, as naturally expected.
However, we observe a critical angle around $25^\circ$ at which the decrease is stopped and the packing fraction reaches
a local minimum. Above this critical angle,
the maximum packing fraction increases with increasing angle and eventually reaches a peak value at $\theta\approx 30^\circ$ before decreasing again.  
The appearance of the local minimum coincides with the emergence of the supported flow regime with a dense core floating on a gaseous layer.
The increase of the packing fraction can be interpreted as the signature of the clustering instability in granular gas \cite{Zanetti1993}.
Importantly, the local maximum of the packing fraction is reached just before the transition towards the asymmetric core regime, as already
noticed for the vorticity. After the local maximum, the packing fraction starts a new decrease with increasing angle. This decrease is concomitant 
with a shrinkage of the latter: particles from the core evaporate and enter the surrounding gaseous region.
with the emergence
of an asymmetric core (as discussed further below) and also 

When we increase the particle hold-up, we obtain the same trend for the packing fraction. The packing fractions at the 
local minima and maxima both increase with increasing particle hold-up but the difference between the maximum and minimum packing fraction
tends to decrease.
We can note that this behavior of the packing fraction is reminiscent of the liquid-gas first-order transition of a molecular gas. 
There is indeed a striking resemblance with the isothermal curves of a simple gas in the pressure-volume diagram.

\begin{figure}[hbt]
\begin{center}
\subfigure[]{\includegraphics[width=0.35\columnwidth]{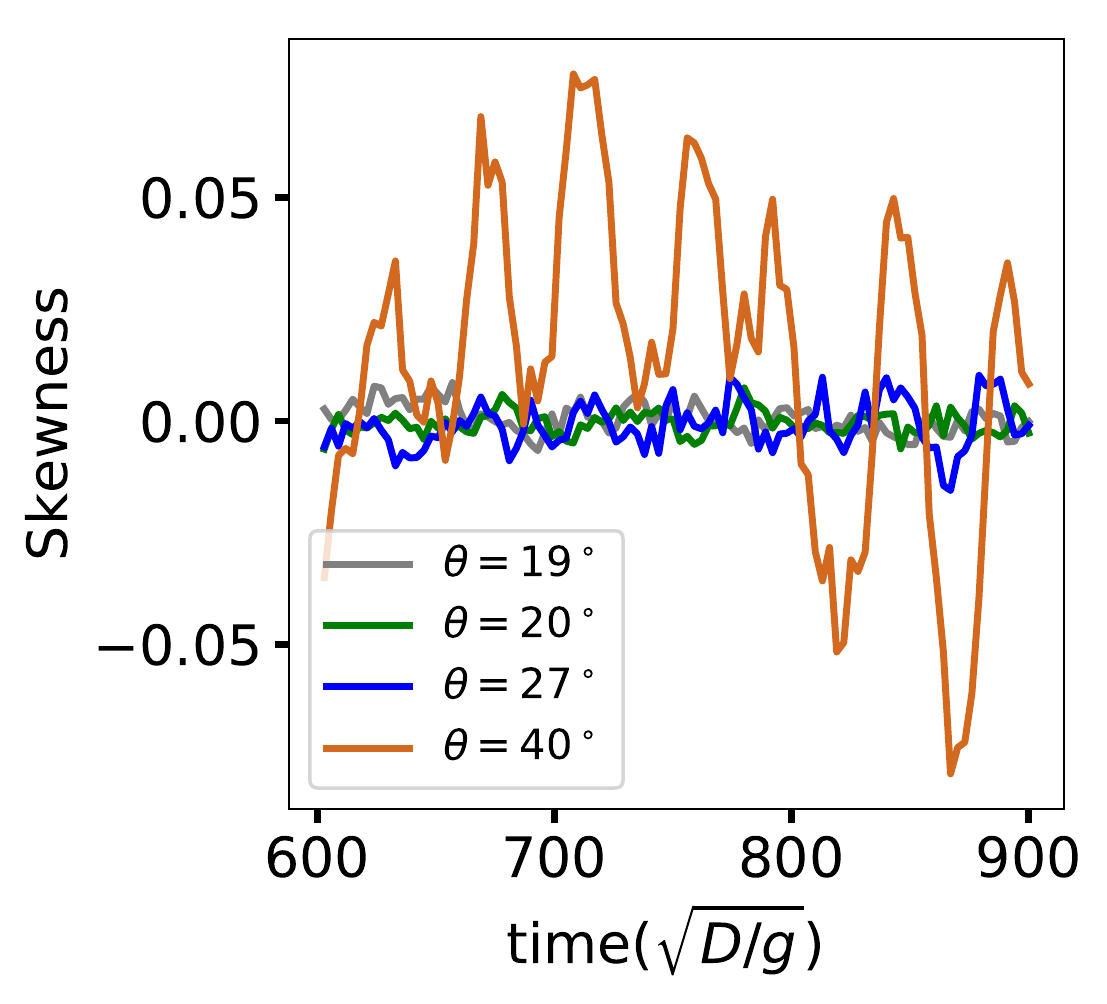}}
\subfigure[]{\includegraphics[width=0.3\columnwidth]{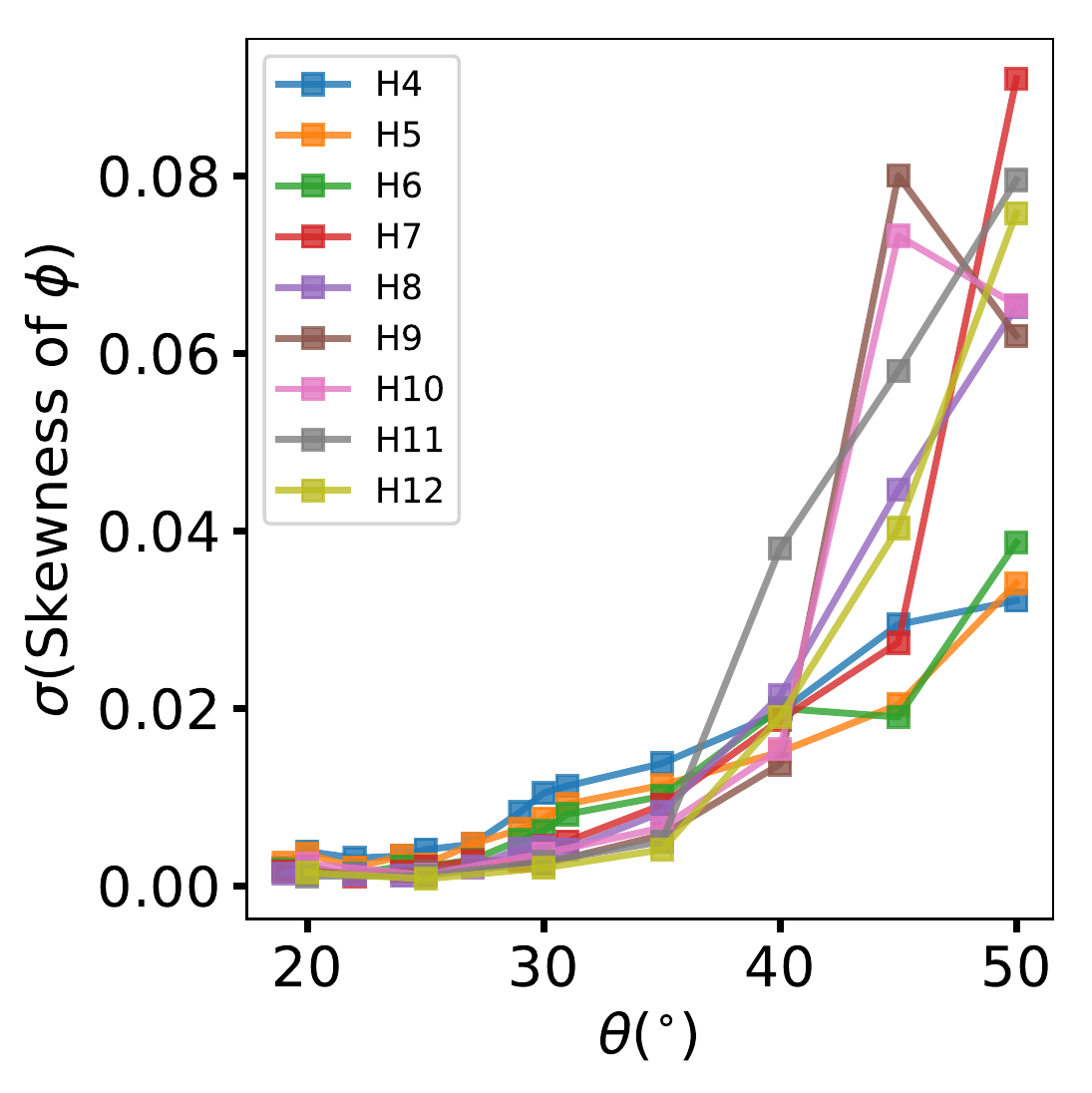}}
\end{center}
\caption{ (a) Evolution of the skewness $S$ of the depth-integrated transverse packing fraction profile $\phi(y)$
as function of time for inclination angles $\theta=19,20,27$ and $40^\circ$ and a fixed particle hold-up $H=6D$.
(b) Standard deviation of the skewness $S$ as a function of the inclination angle for various particle hold-up.
}
\label{asym}
\end{figure}

The last transition concerns the supported regime with a asymmetric core. 
We attempted to characterize the asymmetry of the dense core by investigating the asymmetry of the instantaneous depth-integrated transverse packing 
fraction profiles $\phi(y)$ through the skewness parameter $S$ defined as 
\begin{equation}
S= \frac{\int_0^W dy \,\phi(y)\,(y-\mu)^3/ \left( \int_0^W dy \, \phi(y) \right)
 }{ \left[ \int_0^W dy\, \phi(y)\, (y-\mu)^2 / \left( \int_0^W dy \, \phi(y) \right) \right]^{3/2}}
\end{equation}
with $\mu=\int_0^W dy \,\phi(y)\,y / \int_0^W dy \, \phi(y)$.
We present in Fig.~\ref{asym} the skewness as a function of time and the amplitude of its variation as a function of the inclination angle.
The skewness $S$ remains small for moderate inclination (i.e., $\theta < 30^\circ$) but becomes significant at large inclination angles 
(i.e., $\theta \ge 30^\circ$) and oscillates with a well defined periodicity which is directly related to the rocking motion of the dense core.  
This parameter thus allows to delineate a clear transition
between supported regimes with a symmetric and asymmetric core, respectively. This transition occurs at  $\theta\approx 30^\circ$ for
moderate particle hold-up and seems to increase up to $\theta=35^\circ$ for large particle hold-up (i.e. $H\ge 10D$).

\section{Sidewall and basal friction}
In these types of confined flows, boundaries play an important role in the flow structure. It is thus instructive to investigate in
particular how effective sidewall and bottom friction, defined as the ratio of tangential to normal stresses,
evolve according to the flow regimes reported below. 
Brodu and co-workers \cite{Brodu2015}
showed the sidewall and bottom friction both increase with increasing inclination angle but surprisingly decrease with increasing
particle hold-up. Here, we are going further by investigating how these trends could be cast into simple laws.

\begin{figure}[htb]
\begin{center}
\includegraphics[width=0.4\columnwidth]{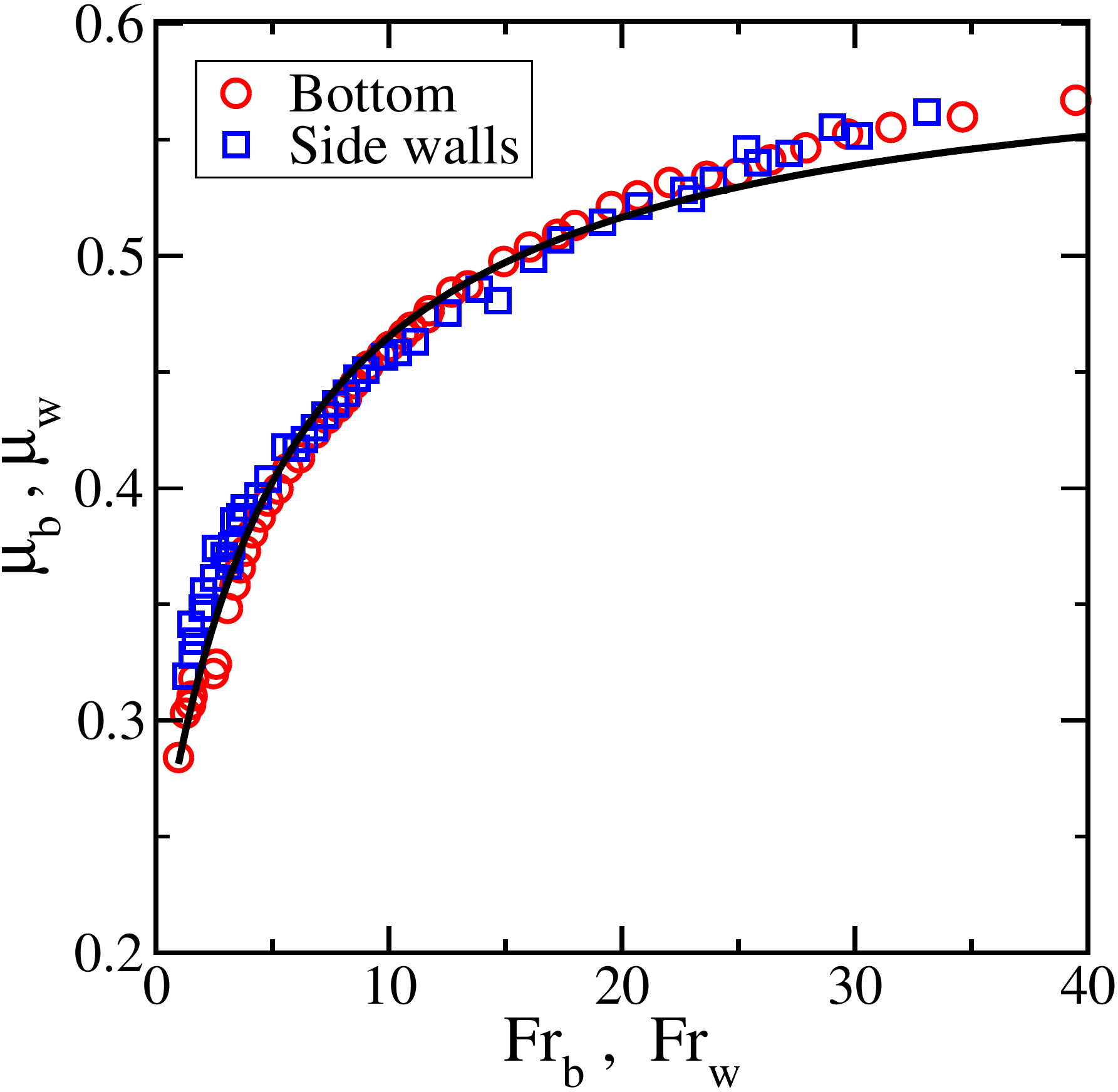}
\end{center}
\caption{ Effective basal friction $\mu_b$ (circle symbols) and sidewall friction $\mu_w$ (square symbols) as a function of the Froude number
$Fr_b=V_{b}/\sqrt{gH \cos \theta}$
and $Fr_w=V_{w}/\sqrt{gH \cos \theta}$, respectively, for all the SFD flow regimes investigated so far 
(i.e., within the parameter range: $4\le H/D \le 12$ and $15^\circ \le \theta \le 50^\circ$).
All the data collapse on a  unique master curve (solid line)  
which is obtained by a fit using Eq.~\ref{fit}.
}
\label{friction}
\end{figure}
We found that the variation of both the basal and sidewall friction can be simply described through a unique dimensionless
number, analog to a Froude number, $Fr=V_{boundary}/\sqrt{gH\cos \theta}$, where $H$ is the particle hold-up, $\theta$ the angle of inclination
and $V_{boundary}$ the velocity at the considered boundary (i.e., either $V_{b}$ or $V_{w}$). Indeed, if we plot the effective basal friction
and sidewall friction as a function of the Froude number $Fr$ for all the SFD flow regimes investigated so far
(i.e., within the parameter range: $4\le H/D \le 12$ and $15^\circ \le \theta \le 50^\circ$).
we get a nice collapse of all the data onto a unique curve (see Fig.~\ref{friction}). 
The $\mu(Fr)$ curve increases monotonically with the Froude number and seems to saturate at large Froude number to a constant asymptotic value,
which is close to the microscopic friction coefficient at the walls, $\mu_g=0.593$. Interestingly, The $\mu (Fr)$ curve shares strong resemblance with
the $\mu(I)$ rheological curve for dense granular flows and  can be well approximated by a similar functional form:
\begin{equation}
\mu(Fr)= \mu_{min} + \frac{\mu_g-\mu_{min}}{1+Fr_0/Fr}
\label{fit}
\end{equation}
with $\mu_{min}\approx 0.22$, and $Fr_0 \approx 5.46$. 

Several comments follow.
It is first important to note that the $\mu (Fr)$ curve 
describes the evolution of the effective friction at the boundaries but not in the 
bulk flow. It can be seen as a boundary condition which can be useful for theoretical approaches. Interestingly, we may wonder
whether the relationship between the effective friction at the walls and the Froude number can be extrapolated to an arbitrary layer 
within the bulk flow.
Second, the $\mu(Fr)$ curve provides a simple explanation for the decrease of the bottom and wall friction with increasing particle hold-up.
Indeed, recalling that the velocities at the boundaries are almost invariant with the particle hold-up (see Eq.~\ref{vbw_eq}),
the Froude number decreases with increasing particle hold-up at a fixed inclination angle. This results in a decrease of the basal friction since
$\mu(Fr)$ is an increasing function of the Froude number. Third, Eq.~\ref{fit} together with Eq.~\ref{vbw_eq} and the definition of the Froude number 
provides us with an explicit expression of the basal and sidewall friction as a function of the inclination angle and particle hold-up.
Fourth, in kinetic theories for granular flows, the effective friction at bumpy wall is often expressed
as a function of the dimensionless quantity $V/\sqrt{T}$ \cite{Jenkins1992}. In the case of flat frictional wall \cite{Jenkins1992}, the relevant quantity
is $g/\sqrt{T}$ where $g=||\vec{V}-(D/2)\vec{\omega} \times \vec{n} ||$ is the contact slip velocity at the wall 
($\vec{n}$ is the unit vector normal to the wall and $\vec{\omega}$ is the mean angular velocity). 
It is thus instructive to check whether the friction
at the basal and side walls can be also described in terms of the ratio $g/\sqrt{T}$. 

\begin{figure}[htb]
\begin{center}
\subfigure[]{\includegraphics[width=0.35\columnwidth]{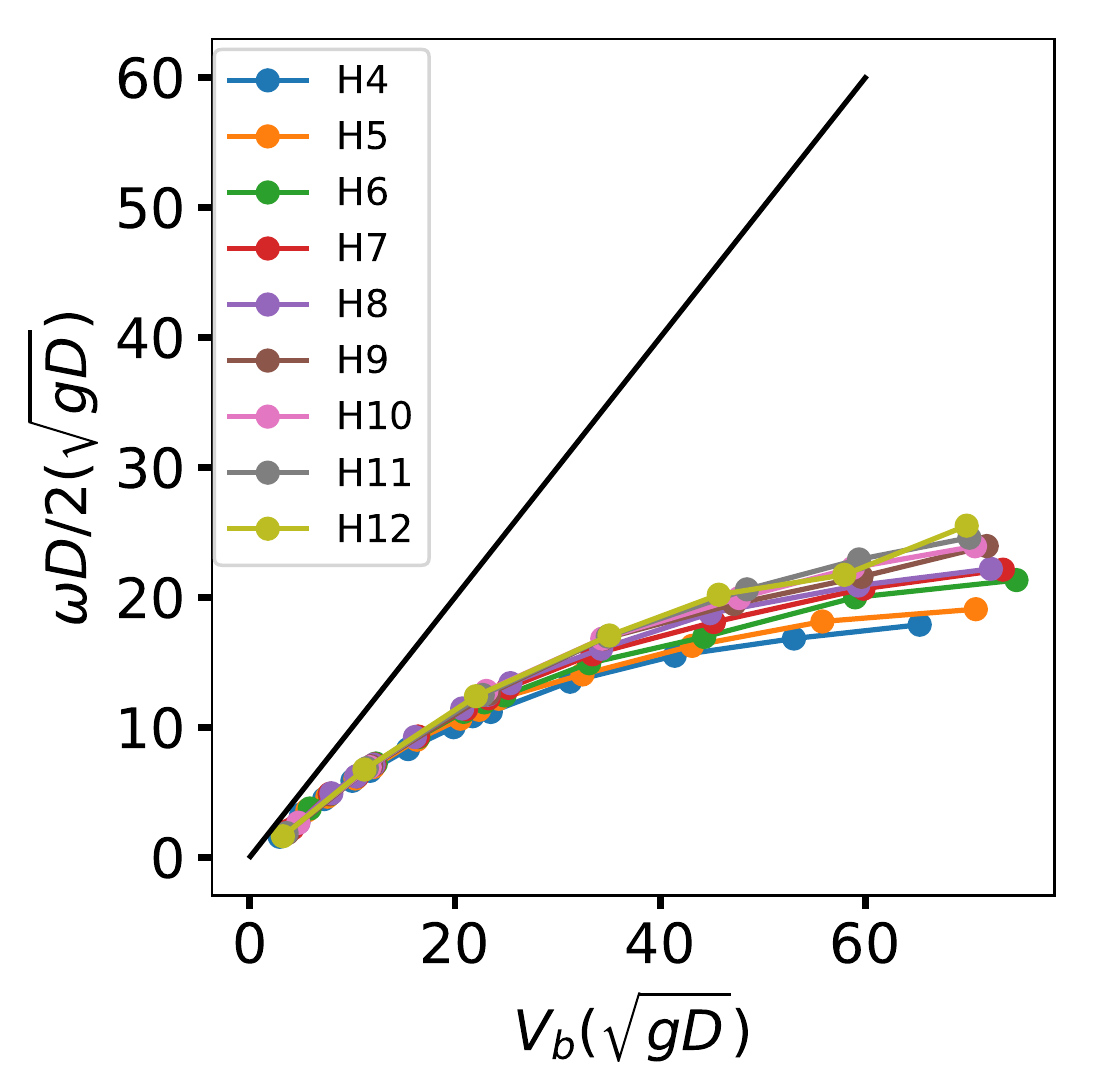}}
\subfigure[]{\includegraphics[width=0.35\columnwidth]{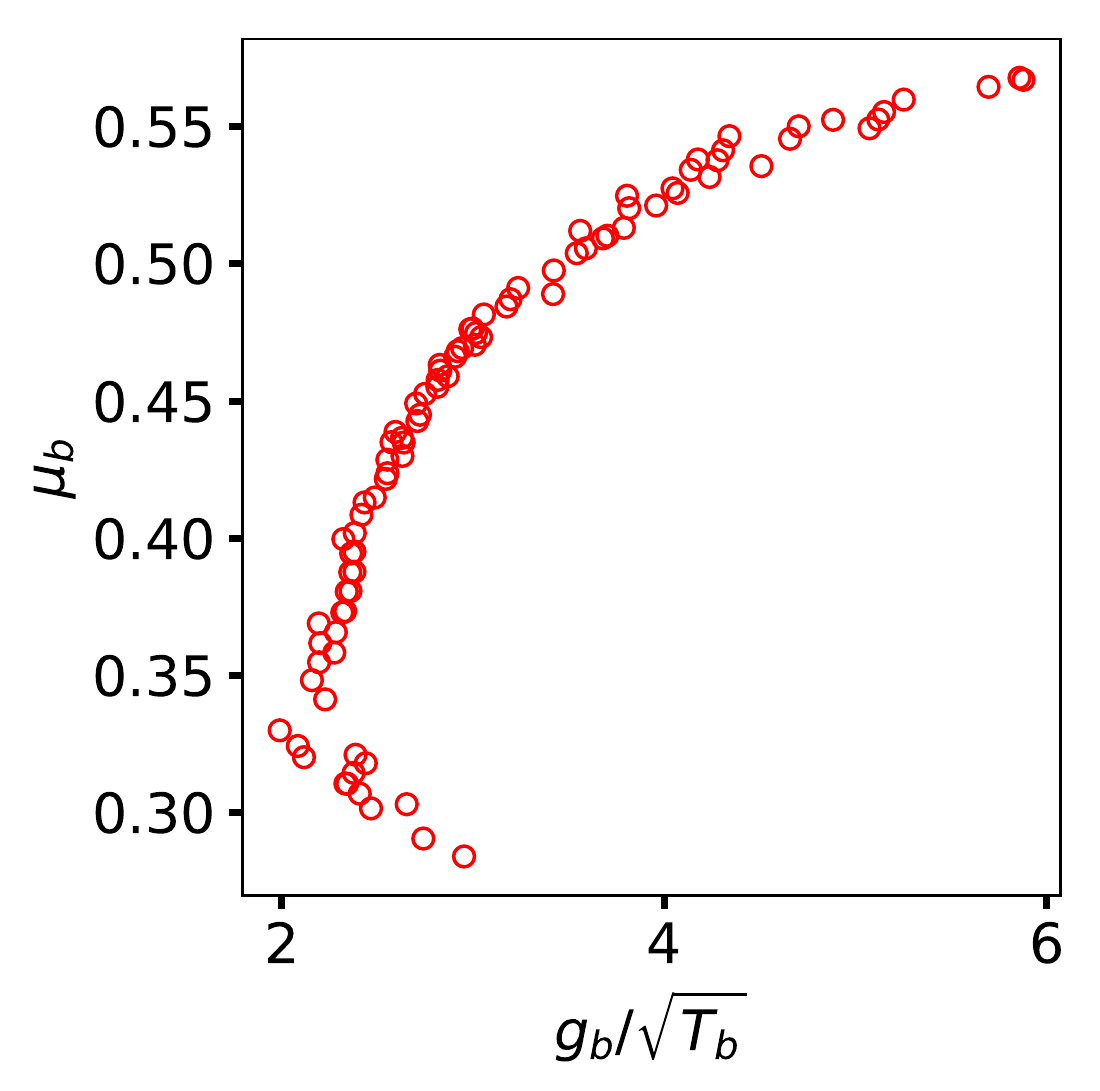}}
\end{center}
\caption{ (a) Angular velocity $\omega$ as a function of the particle velocity $V_b$
at the bottom for increasing particle hold-up from $H=4D$ to $H=12D$. We used the same color code as in Fig.~10 and 11.
(b) Bottom friction $\mu_b$ as a function of the dimensionless contact slip velocity $g_b/\sqrt{T_b}$.
}
\label{mub_vs_gb}
\end{figure}

We show in Fig.~\ref{mub_vs_gb}.a 
the angular speed of the particle at the bottom multiplied by the particle radius, i.e, $w(D/2)$, as function of
the particle velocity $V_b$ at the bottom. We can note that for small angle (i.e.,$\theta \le 20^\circ$ or equivalently $V_b\le 5$), 
the rotation speed $w(D/2)$ is very close to the particle velocity indicating that the particles roll without sliding. 
At higher angle particle, this is no more the case: the 
particles thus roll with sliding. With this, we can compute the slip velocity at the bottom, $g_b$, and plot the effective bottom friction as a function of 
the dimensionless quantity $g_b/\sqrt{T_b}$ (see Fig.~\ref{mub_vs_gb}.b).
We find a nice collapse of the data on a single curve which is very similar to the $\mu(Fr)$ curve. We can note however
a deviation of the monotone behavior at low value of the friction (i.e., for small inclination angles corresponding to dense flows).
It thus turns out that the Froude number and the dimensionless contact slip velocity $g/\sqrt{T}$ play a similar role
and are closely related. We find indeed the following correlation:
\begin{equation}
Fr_{b} \approx 9.95 \left( \frac{g_{b}}{\sqrt{T_b}} -1.86 \right)
\end{equation}
The above correlation works well for large Froude number but fails for small Froude number below $2$ corresponding to dense flows.

\section{Conclusion}
We have studied high-speed confined granular flows down inclines and describe in detail the different SFD flow regimes,
including the supported regime which display striking properties. We have highlighted that the friction at the basal
and side walls can be described by a unique curve that depends solely of the Froude number defined
as $V/\sqrt{gH\cos\theta}$, where $V$ is the particle velocity at the walls.

A crucial question is the extent to which the supported regimes and their features are
specific to the material parameters and the confined geometry that we have considered.
Additional simulations where the material parameters (friction and
restitution coefficient) and  confinement $W$ are varied are going. Preliminary results
show that supported flows are very robust to parameter change but their onset of appearance may be
drastically affected. For example, increasing the dissipation in the grain-grain collision favors
the development of supported flows.

Finally, these results provide a unique set of very complex granular flow regimes for
testing theoretical and rheological models. \\

\noindent{\bf Acknowledgements:}
We acknowledge the support of the French Research National Agency through the project ANR-16-CE01-0005. 


\end{document}